\documentclass[useAMS,preprint]{nature_simon}
\usepackage{color}

\usepackage{amssymb}
\usepackage{aas_macros}
\usepackage{amsfonts}
\usepackage{amsmath}
\usepackage{geometry}
\usepackage[pdftex]{graphicx}               %for importing graphics

\def\apjl{{Astrophys. J.}}                % Astrophysical Journal, Letters
\def\apjs{{ApJS}}               % Astrophysical Journal, Supplement
\def\ao{{Appl.~Opt.}}           % Applied Optics
\def\aap{{Astron. Astrophys.}}                % Astronomy and Astrophysics
\def\ao{Appl. Opt.}

\def\nt/f{Nucl. Tech./Fusion}

\def\ell{Electron. Lett.}

\title{Observations of gas flows inside a protoplanetary gap}

\author{Simon Casassus$^{1}$,
Gerrit van der Plas$^{1}$,
Sebastian Perez~M.$^{1}$,
William R.F.Dent$^{2,3}$,
Ed Fomalont$^4$,
Janis Hagelberg$^5$,
Antonio Hales$^{2,4}$,
Andr\'es Jord\'an$^{6}$,
Dimitri Mawet$^3$,
Francois M\'enard$^{7,8}$,
Al Wootten$^4$,
David Wilner$^{9}$,
A.~Meredith Hughes $^{10}$,
Matthias R. Schreiber$^{11}$,
Julien H. Girard$^3$,
Barbara Ercolano$^{12}$
Hector Canovas$^{11}$,
Pablo E. Rom\'an$^{13}$,
Vachail Salinas$^1$
}

\begin{document}

\maketitle

\begin{enumerate}
 \item Departamento de Astronom\'{\i}a, Universidad de Chile, Casilla 36-D, Santiago, Chile
% \item Millenium Nucleus ``Protoplanetary Disks in ALMA Early Science''
 \item Joint ALMA Observatory, Alonso de C\'ordova 3107, Vitacura 763-0355, Santiago - Chile
 \item European Southern Observatory (ESO), Casilla 19001, Vitacura, Santiago, Chile
 \item National Radio Astronomy Observatory, 520 Edgemont Road, Charlottesville, VA 22903-2475, USA
 \item Observatoire de Gen\`eve, Universit\'e de Gen\`eve, 51 ch. des Maillettes, 1290, Versoix, Switzerland
 \item Departamento de Astronom\'{\i}a y Astrof\'isica, Pontificia  Universidad Cat\'olica de Chile, Santiago, Chile
 \item UMI-FCA, CNRS / INSU France (UMI 3386) , and Departamento de Astronom\'{\i}a, Universidad de Chile, Santiago, Chile.
 \item  CNRS / UJF Grenoble 1, UMR 5274, Institut de Plan\'etologie et d’Astrophysique de Grenoble (IPAG), France
 \item Harvard-Smithsonian Center for Astrophysics, 60 Garden Street, Cambridge, MA 02138 USA
 \item Department of Astronomy, U. C. Berkeley, 601 Campbell Hall, Berkeley, CA 94720
 \item Departamento de F\'{\i}sica y Astronom\'{\i}a, Universidad Valparaiso, Av. Gran Bretana 111, Valparaiso, Chile.
 \item University Observatory, Ludwig-Maximillians University, Munich.
 \item Center of Mathematical Modeling, University of Chile, Av. Blanco Encalada 2120 Piso 7, Santiago, Chile.
\end{enumerate}

\begin{abstract}

%Giant planet formation is thought to occur in the first few million
%years following stellar birth. Models8 demonstrate that giant planet
%formation carves a deep gap in the dust component (shallower in the
%gas9-11). Infrared observations of the disk around the young star HD
%142527 found an inner disk ∼10 AU in radius1, surrounded by a
%particularly large gap, with an outer disk beyond 140 AU2, indicative
%of a perturbing planetary-mass body at ∼ 90 AU.  Molecular gas4,5 is
%distributed with a horseshoe morphology coincident with the IR outer
%disk4.  The vigorous stellar accretion rate6 would deplete the inner
%disk7 in less than a year, so in order to sustain the observed
%accretion, matter must flow from the outer-disk into the cavity and
%cross the gap.
%
%Here we report observations with the Atacama Large Millimetre
%Array (ALMA) that reveal diffuse CO gas inside the gap, with denser
%HCO+ gas along gap-crossing filaments, and confirm the horseshoe
%morphology of the outer disk. The estimated flow rate of the gas is
%..., which is sufficient to maintain accretion onto the star at the
%present rate for ... years.
%

%with HCO+, tracing denser gas, along gap-crossing... 

Gaseous giant planet formation is thought to occur in the first few
million years following stellar
birth. Models\cite{Lubow2006ApJ...641..526L} predict that the process
carves a deep gap in the dust component (shallower in the
gas\cite{Fouchet2010,Ayliffe2012,Zhu2011}).  Infrared observations of
the disk around the young star HD~142527, at $\sim$140~pc, found an
inner disk $\sim$10~AU in radius\cite{2004Natur.432..479V}, surrounded
by a particularly large gap\cite{Fukagawa2006}, with a
disrupted\cite{Casassus2012ApJ...754L..31C} outer disk beyond 140~AU,
indicative of a perturbing planetary-mass body at $\sim$90~AU.  From
radio observations\cite{2008Ap&SS.313..101O, 2011ApJ...734...98O} the
bulk mass is molecular and lies in the outer disk, whose continuum
emission has a horseshoe morphology\cite{2008Ap&SS.313..101O}. The
vigorous stellar accretion rate\cite{2006A&A...459..837G} would
deplete the inner disk\cite{2011A&A...528A..91V} in less than a year,
so in order to sustain the observed accretion, matter must flow from
the outer-disk into the cavity and cross the gap. In dynamical models,
the putative protoplanets channel outer-disk material into
gap-crossing bridges that feed stellar accretion through the inner
disk\cite{2011ApJ...738..131D}. Here we report observations with the
Atacama Large Millimetre Array (ALMA) that reveal diffuse CO gas
inside the gap, with denser HCO$^+$ gas along gap-crossing filaments,
and that confirm the horseshoe morphology of the outer disk. The
estimated flow rate of the gas is in the range $7~10^{-9}$ to
$2~10^{-7}$~M$_\odot$~yr$^{-1}$, which is sufficient to maintain
accretion onto the star at the present rate.

\end{abstract}

%However, the vigorous stellar accretion rate\cite{2006A&A...459..837G}
%of $7 \times 10^{-8}$~M$_{\odot}$~yr~$^{-1}$, would quickly deplete
%the $10^{-9}$~M$_{\odot}$ inner disk dust
%mass\cite{2011A&A...528A..91V}, for any physical gas to dust mass
%ratios.

%The ALMA data show three important features of this disk that
%we summarize in Fig.~\ref{fig:cycle0}: [1] Diffuse CO-emitting gas
%inside the disk gap; [2] Gap-crossing flows in the dense gas tracer
%HCO+; [3] A “horseshoe” shape of the continuum emission.

%Hydrodynamical simulations of a 10~M$_\mathrm{jup}$ body in a circular
%orbit at $r = 90$~AU evolve through transient configurations that
%qualitatively match the
%observations\cite{Casassus2012ApJ...754L..31C}.

%Non-Keplerian, high-velocity gaseous flows in
%the high gas density tracer HCO$^+$ near the star; [3] Gap-crossing
%bridges in systemic velocity HCO$^+$;

% in amounts that increase with decreasing companion mass

% A ``horseshoe'' shape of
%the continuum emission.

%\textbf{1: Diffuse CO gas inside the disk gap:} 

The HD~142527 system offers an unhindered view of its large central
cavity, and is a promising laboratory to observe on-going gaseous
giant planet formation.  There is good understanding of the
orientation of the disk. Multi-wavelength data are consistent with an
inclination of $\sim$20$^{\circ}$, so almost
face-on\cite{2011A&A...528A..91V}. The disk position angle is
$\sim -20$~deg east of north and the eastern side is the far side
of the disk, as suggested by a clear view of the outer disk's inner
rim in the mid-IR\cite{Fujiwara2006, 2011A&A...528A..91V} and by a
clock-wise rotation suggested by a probably trailing spiral arm to the
west\cite{Fukagawa2006}.

We find that the CO(3-2)-emitting gas peaks inside the cavity.  Other
disks have been observed to exhibit a CO decrement within dust
cavities\cite{Lyo2011AJ....142..151L, 2012ApJ...753...59M}, and may
represent later evolutionary stages or different gap clearing
mechanisms. Gas inside dust cavities has previously been directly
observed very close to the central star (inside the dust evaporation
radius) using near-IR interferometry\cite{Tatulli2007A&A...464...55T,
  Kraus2008A&A...489.1157K, Eisner2010ApJ...718..774E}. Other indirect
observations of gas inside\cite{Carr2001ApJ...551..454C,
  Najita2003ApJ...589..931N, Acke2006A&A...449..267A,
  vanderPlas2008A&A...485..487V, Salyk2009ApJ...699..330S} dust gaps
at larger distances from the central star have interpreted
spectroscopically resolved gas tracers, such as ro-vibrational
CO~4.67$\mu$m and [O\,{\sc i}]~6300\AA, under the assumption of
azimuthal symmetry and Keplerian rotation.  Spectro-astrometry in
combination with Keplerian disk models and azimuthal symmetry has also
been used to infer the presence of CO gas inside disk
gaps\cite{Pontopiddan2008,vanderPlas2009A&A...500.1137V,
  Pontoppidan2011ApJ...733...84P}. Our new data provide a
well-resolved observation of gas at sub-mm wavelengths inside a dust
cavity.

%Salyk2007ApJ...655L.105S,
%Our observations allow, for the
%first time, to unambiguously trace the structure of gas inside a disk
%gap.

%(a technique that traces the 1-dimensionally projected spatial center
%of mass of an emission line)

%Salyk2007ApJ...655L.105S
% Bagnoli2010ApJ...724L...5B

\begin{figure}
\begin{center}
\includegraphics[width=\textwidth,height=!]{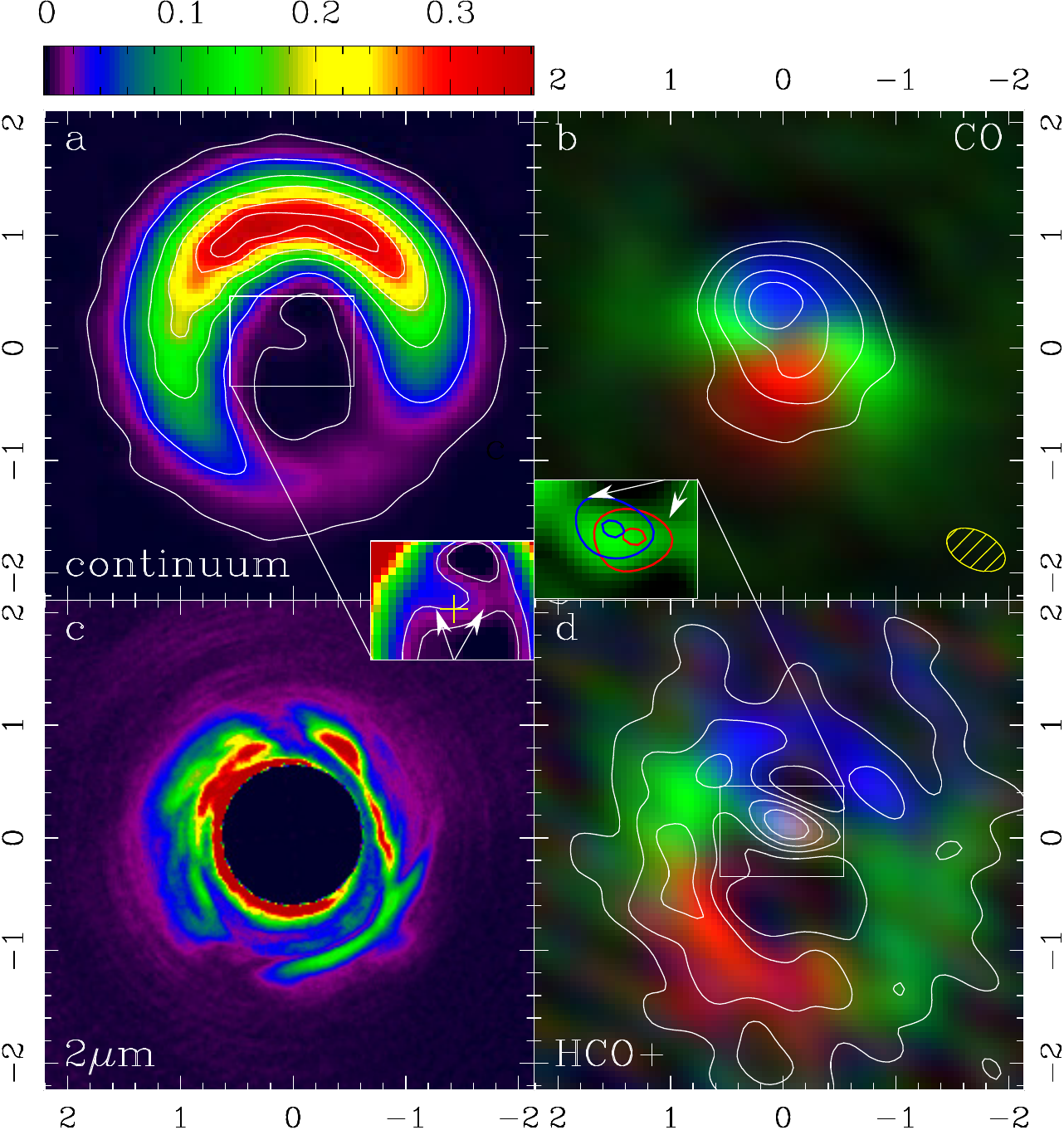}
\end{center}
\caption{{\bf ALMA observations of HD~142527, with a horseshoe dust
    continuum surrounding a cavity that still contains gas.} We see
  diffuse CO gas in Keplerian rotation (coded in doppler-shifted
  colours), and filamentary emission in HCO$^+$, with non-Keplerian
  flows near the star (comparison models illustrative of Keplerian
  rotation are shown in SI). The near-IR emission abuts onto the inner
  rim of the horseshoe-shaped outer disk. The star is at the origin of
  coordinates and axis labels are in arcsec; North is up, East is to
  the left. {\bf a:} The image labelled `continuum' is a CLEANed image
  of the continuum at 345~GHz, with specific intensity units in
  Jy~beam$^{-1}$. It is shown in exponential scale. A beam ellipse is
  shown in the bottom right, and contour levels are taken at 0.01,
  0.1, 0.3, 0.5, 3/4, 0.9 times the peak value. The noise level is
  1~$\sigma$ =0.5~mJy~beam$^{-1}$. {\bf b:} The image labelled 'CO' is
  CO(3-2) line intensity in white contours at fractions of 0.3, 0.5,
  0.75 and 0.95 of the peak intensity value
  $2.325~10^{-20}$~W~beam$^{-1}$.  The underlying RGB image also shows
  CO(3-2) line intensity but integrated in three different velocity
  bands, whose velocity limits are indicated in the spectra of
  Fig.~\ref{fig:spectra}.  {\bf c:} The image labelled '2~$\mu$m' is a
  near-IR image from Gemini that traces reflected stellar light, shown
  in linear scale. We applied an circular intensity mask to the
  stellar glare, some of which immediatelly surrounds the mask. See
  Fig.~\ref{fig:registration} in SI for an overlay with the
  continuum. {\bf d:} The image labelled 'HCO$^+$' shows HCO$^+$(4-3)
  line intensity in white contours at fractions of 0.1, 0.3, 0.5, 0.75
  and 0.95 of the peak intensity value $0.40~10^{-20}$~W~beam$^{-1}$,
  overlaid on an RGB image of HCO$^+$ intensity summed in three
  different colour bands, with definitions in Fig.~\ref{fig:spectra}.
  {\bf Insets:} Zoom on the central features that cross the dust
  gap. The cross indicates the star at the origin, with a 0.05~arcsec
  accuracy, and the arrows point at the filaments.
  Fig.~\ref{fig:cycle0}\,a inset: same as Fig.~\ref{fig:cycle0}\,a,
  with a narrow exponential scale highlighting the gap-crossing
  filaments.  Note how these features appear to grow from the eastern
  and western sides of the horseshoe.  Contours are at at 0.0015 and
  0.005~Jy~beam$^{-1}$.  Fig.~\ref{fig:cycle0}\,d inset: deconvolved
  models (see SI) of the HCO$^+$ emission at velocities where the
  gap-crossing filaments are seen, i.e. from 3.2 to 4.3~km~s$^{-1}$,
  in green. Intensity maps for the blue and red velocity ranges (see
  Fig.~\ref{fig:spectra} for definitions) are shown in contours, with
  levels at 0.5 and 0.95 of the peak values. These red and blue
  contours are an alternative way to present the intensity field shown
  in Fig.~\ref{fig:cycle0}\,d, but deconvolved for ease of
  visualization.\label{fig:cycle0} }
\end{figure}

The dust gap that we see in radio continuum (Fig.~\ref{fig:cycle0}\,a)
indicates a drop by a factor of at least 300 in surface density of
mm-sized grains, from the contrast ratio between the peak on the
horseshoe-shaped outer disk (the northern peak at
360~mJy~beam$^{-1}$), and the faintest detected signal inside the gap
(namely the western filament at 1~mJy~beam$^{-1}$, see below). Yet
there is no counterpart in the CO(3-2) map (Fig.~\ref{fig:cycle0}\,b)
of the arc that we see in continuum. CO(3-2) is likely optically
thick, as reflected by its diffuse morphology, so it traces the
temperature profile rather than the underlying density field.
Exhaustive modelling of optically thin isotopologue data is required
to accurately constraint the depth of the gaseous gap. To study the
distribution of dense gas inside the cavity we use the high density
gas tracer HCO+.

%In
%our comparison line transfer models a gaseous gap depth of 100 shows a
%decrement inside the ring.

% A mere $10^{-9}~M_\odot$ of CO mass could account for the observed
% $^{12}$CO(3-2) flux.

%We also observed in HCO$^+$, a tracer of
%dense gas, unraveling the structure of the gas inside the cavity.

%Previous observations of gas inside dust cavities through optical/IR
%spectroscopy have relied on model kinematic velocity
%fields\cite{Pontopiddan2008}.

%\textbf{2: Gap crossing bridges in the high gas density tracer HCO$^+$
%  and in continuum}:

%, i.e. at 2.7 and 3.8~km~s$^{-1}$ in 10-channel-averaged bins

The second result from our observations is that the HCO$^+$(4-3)-emitting
gas, expected in the denser regions ($n_{\mathrm{H}2} \sim
10^{6}$~cm$^{-3}$) exposed to UV radiation, is indeed found in the
exposed rim of the dense outer disk, but also along gap-crossing
filaments. The most conspicuous filament extends eastwards from the
star, while a fainter filament extends westwards. Both filaments
subtend an angle of $\sim 140\pm10$~deg with the star at its
vertex. The central regions of these filaments correspond to the
brightest features in the HCO$^+$ line intensity maps
(Fig~\ref{fig:cycle0}~c), although the outer disk is brighter in peak
specific intensity (SI, Fig~\ref{fig:whcomax}).  Thus line velocity
profiles are broader in the stellar side of the filaments than in the
outer disk, where they merge with the outer disk Keplerian rotation
pattern. These narrow and systemic velocity HCO$^+$ filaments are best
seen in intensity maps integrated over the filament velocities (inset
to Fig.~\ref{fig:cycle0}\,d). No central peak is seen in the channel
maps (SI, Fig.~\ref{fig:channelsHCOplus}), so that a beam-elongation
effect can be ruled out. The eastern filament also stands out in peak
HCO$^+$ specific intensity (SI, Fig.~\ref{fig:whcomax}\,e).  For ease
of visualization we show deconvolved models of the HCO$^+$ intensity
images in the inset to Fig.~\ref{fig:cycle0}\,d.  A related feature is
seen in CO(3-2), whose intensity peaks in the more diffuse regions
surrounding the eastern HCO$^+$ filament. Interestingly we note from
the inset to Fig.~\ref{fig:cycle0}\,a that the continuum also shows
features under the HCO$^+$ filament, faint and growing away from the
walls of the horseshoe-shaped outer-disk. Estimates of physical
conditions are given in SI.

%The continuum intensities of the bridges seem anti-correlated with
%HCO$^+$, they decrease closer to the star.  For constant dust mass
%density we would expect the opposite trend, since the mm-sized grains
%emitting at 345~GHz cary most of the dust mass, and should be brighter
%in the hotter central regions.

The molecular and filamentary flows near the star are non-Keplerian.
Blue-shifted emission extends to the east from the central intensity
peak (Fig.~\ref{fig:cycle0}\,c). This velocity component is broad near
the star, with emission ranging from -3.4 to +11~km~s$^{-1}$ (SI,
Fig.~\ref{fig:channelsHCOplus}), and is marginally resolved (the
central HCO$^+$ peak extends over $\sim 0.65\times0.38$~arcsec). In
the deconvolved images of the inset to Fig.~\ref{fig:cycle0}\,d the
peak intensity in the blue and red are separated by $\sim$0.2~arcsec,
i.e. by a diameter of the inner disk, and at a position angle (PA)
orthogonal to that expected from close-in high-velocity material in
Keplerian rotation. Very blue-shifted emission could reach out to
0.2~arcsec from the star (SI, channel at --2.4~km~s$^{-1}$ in
Fig.~\ref{fig:channelsHCOplus}, taking into account the beam). A
blue-shifted CO(3-2) high-velocity component can also be seen at the
root of this feature, near the star (SI, at --2.1~km~s$^{-1}$ in
Fig.~\ref{fig:channelsCO}).

%Given our
%astrometric accuracy of 0.05~arcsec, v

%p (sqrt(1.2)*5.9*sqrt(140/80))

%Given an inclination of 20~deg, an outflow
%orthogonal to the plane of the disk would lie along the line of
%sight. **** 

%
%*****
%A natural interpration for the filaments discovered by ALMA is
%gap-crossing accretion flows, or `bridges'. The gap seen in HD142527
%can only be carved by multiple planets, this opens the exciting
%possibly that these accretion are induced by planets. 
%*****
%

The non-Keplerian HCO$^+$ is probably not consistent with a central
outflow. Stellar outflows are not
observed\cite{Sacco2012ApJ...747..142S} in disks with inner cavities
and no molecular envelopes (i.e. transition disks). For an outflow
orientation, the low velocities measured by the lines imply that the
filaments in HD~142527 would stand still and hover above the star (SI,
Sec.~\ref{sec:conditions}). Even the blue-shifted emission is slow by
comparison to escape velocity.  A slow disk wind (e.g. photoevaporative or
magnetic driven) can also be excluded on the basis of the high
collimation shown by the HCO+ emission.  Indeed the CO~4.67~$\mu$m
seen in the inner disk\cite{Pontoppidan2011ApJ...733...84P} is purely
Keplerian, it does not bear the signature of the disk winds seen in
other systems, and its orientation is common to the outer disk.  An
orthogonal inner disk can be also be discarded on dynamical grounds
(SI, Sec.~\ref{sec:conditions}).

It is natural to interpret the filaments as planet-induced
gap-crossing accretion flows, or `bridges'.  Since the eastern side is
the far side, the blue-shifted part of the eastern bridge is directed
towards the star, and is a high-velocity termination of the accretion
flow onto the inner-disk. These bridges are predicted by
hydrodynamical simulations when applied to planet-formation feedback
in HD~142527\cite{Casassus2012ApJ...754L..31C}.  In these simulations
the bridges straddle the protoplanets responsible for the dynamical
clearing of the large gap in HD~142527. They are close to Keplerian
rotation in azimuth, but do have radial velocity components at
$\gtrsim 1/10$ of the azimuthal components. In our data we see that as
the bridges land onto the inner disk they also coincide with higher
velocity material, at (2-D) radial velocities inferior to their
azimuthal velocities in the plane of the sky.

An interesting comparison object is GG~Tau. A sub-mm continuum
accretion stream\cite{Pietu2011A&A...528A..81P} is seen to cross the
gap in this circum-binary disk, with indications of shocked IR
molecular gas in the inner disk\cite{Beck2012ApJ...754...72B}. The
angular radii of the rings in GG~Tau and in HD~142527 are very
similar, as are the morphologies of the GG~Tau streamer and the
eastern filament (although it is fainter relative to the outer disk in
HD~142527). However, the GG~Tau binary has a mass ratio $\sim$1 with a
separation of 0.25~arcsec, and is aligned along the streamer, while in
HD~142527 no stellar companion has been detected (see SI for limits,
mass ratios are $> 10$ at 0.088~arcsec).  The putative companions
responsible for the streams in HD~142527 are much lower mass than in
GG~Tau.

We performed high-contrast IR imaging to attempt the detection of the
possible accreting protoplanets that would be expected if the gap
crossing bridges observed in HCO$^+$ are indeed planet-induced
gap-crossing accretion flows. Neglecting extinction, we could
virtually rule out any companion more massive than
$\sim$~4~M$_\mathrm{jup}$ at separations from 0.3 to 2.5~arcsec (SI,
Sec.~\ref{sec:ADI} and Fig.~\ref{fig:ADI}, also on the lack of close
stellar companions).  However, according to the hydrodynamical
simulations the channeling protoplanets should be located inside the
gap-crossing bridges.  Our estimates for $N_H$ along the bridges
correspond to a broad range of high extinction values $5 \lesssim A_V
\lesssim 50$, for standard dust abundances. Any protoplanets embedded
inside the bridges, and certainly those embedded in the dense
horseshoe structure, will be obscured, and our mass limits will be
correspondingly increased.

A third feature of our observations is the horseshoe shape of the
continuum, seen previously by the Sub-Millimetre
Array\cite{2008Ap&SS.313..101O} at coarser resolutions, but whose
origin is still unclear.  The mm continuum traces the total dust mass,
so the north-south specific intensity ratio of $28\pm0.5$ reflects the
underlying dust mass asymmetry. At its peak the continuum may even be
optically thick, as it coincides with a decrease in the HCO+
emission. For a constant gas-to-dust mass ratio, such horseshoe-shaped
mass asymmetries arise in models of planet-induced dynamical
clearing. In general these horseshoes can be produced by Rossby wave
instabilities, which are seen in high-resolution 3D simulations at the
edge of sharp density gradients (Varni\`{e}re private
communication). However, horseshoes have also been modelled in the
context of large-scale vortices induced by sharp viscosity
gradients\cite{2012MNRAS.419.1701R}.

Another interpretation for the horseshoe continuum is varying
dust-to-gas ratio and azimuthal grain-size segregation.  By contrast
to the continuum, the outer disk is seen as a whole ring in HCO$^+$
(Fig.~\ref{fig:whcomax}), which is a tracer of dense gas, and is
probably optically thick along the ring.  A rarefaction of mm-sized
dust grains to the south could perhaps explain the lack of 345~GHz
signal. Only small dust grains would be found in the south. These
small grains efficiently scatter the near-IR light seen in
Fig.~\ref{fig:cycle0}\,c, filling the opening of the
horseshoe. However, azimuthal dust segregation has been predicted for
centimetre-sized grains at co-rotation in two-fluid simulations of
cavity clearing by giant planet formation\cite{Fouchet2010}, while
mm-sized grains remain relatively unaffected. We explain in SI why
differential stellar heating cannot account for the observed
north-south contrast.

% (Pinilla et al. 2012, A&A, 545)

The filamentary flows and the residual gas inside the cavity are in
qualitative agreement with planet formation feedback on the parent
disk, that carves a gap in the dust distribution while still feeding
stellar accretion through gap-crossing accretion streams. As detailed
in SI, the observed inflow velocity, together with the critical
density of the molecular tracer and the section of the filaments,
provide a lower bound to the mass inflow rate of
$7~10^{-9}~$M$_\odot$~yr$^{-1}$.  An upper bound of
$2~10^{-7}~$M$_\odot$~yr$^{-1}$ can be estimated from the continuum mass in
the filaments and their kinematic timescale. These estimates for the
mass inflow rate are close to the observed stellar accretion
rate\cite{2006A&A...459..837G} of $7 \times
10^{-8}$~M$_{\odot}$~yr~$^{-1}$, bringing quantitative support to our
suggestion that the HCO$^+$ filaments are inflows.

%This initial result on protoplanetary disks will be followed by more
%detailed observations as the ALMA array grows and reaches completion.

%The dotted yellow line corresponds to the half-maximum intensity level
%in a deconvolved model (see SI) of the continuum shown in `a'.

%This image is
%  integrated over a velocity range that is common to the green range
%  of Fig.~\ref{fig:cycle0}\,b and Fig.~\ref{fig:cycle0}\,d.

%{\bf b:} HCO$^+$(4-3) line
%  intensity outlining similar bridges as in continuum but with a
%  bright center near the star. Contours are shown at 3$\sigma$ and
%  5$\sigma$ where $1\sigma= 0.028~10^{-20}$~W~beam$^{-1}$.  

%% Put the bibliography here, most people will use BiBTeX in
%% which case the environment below should be replaced with
%% the \bibliography{} command.

%\bibliography{/Users/simon/common/texinputs/merged.bib}

%\begin{thebibliography}{1}
%\bibitem{dummy} Articles are restricted to 50 references, Letters
%to 30.
%\bibitem{dummyb} No compound references -- only one source per
%reference.
%\end{thebibliography}
%

%% Here is the endmatter stuff: Supplementary Info, etc.
%% Use \item's to separate, default label is "Acknowledgements"

\begin{addendum}

\item [Supplementary Information] additional figures, information
  on instrumental setups and physical condition estimates. 

%is linked to the online version of the paper. 

\item This paper makes use of the following ALMA data:\\ {\tt
  ADS/JAO.ALMA\#2011.0.00465.S}. ALMA is a partnership of ESO, NSF,
  NINS, NRC, NSC, ASIAA. The Joint ALMA Observatory is operated by
  ESO, AUI/NRAO and NAOJ. Also based on observations obtained at the
  Gemini Observatory. Financial support was provided by Milenium
  Nucleus P10-022-F (Chilean Ministry of Economy), and additionally by
  grants FONDECYT 1100221 and Grant 284405 from the European Union FP7
  Programme.

 \item[Contributions] General design of ALMA project, data analysis
   and write-up: SC.  Discussion of IR observations of gas in
   cavities: GvdP.  Hydrodynamical modeling: SP. ALMA data reduction:
   AH, EF.  ADI processing: DM, JH, JG. Contributions to ALMA Cycle~0
   proposal: AJ, FM, DW, MH.  Design of ALMA observations: AW, AH,
   SC. Authors WD to AW contributed equally.  All authors discussed
   the results and commented on the manuscript.

%SD processing: SC, AJ, VS.

\item[Competing Interests] The authors declare that they have no
competing financial interests.

 \item[Correspondence] Correspondence and requests for materials
should be addressed to S.C.~(email: scasassus@u.uchile.cl).
\end{addendum}

%%
%% TABLES
%%
%% If there are any tables, put them here.
%%

\newpage

\setcounter{page}{1}
\setcounter{figure}{0}
\renewcommand{\thefigure}{S\arabic{figure}}

\begin{flushleft}
{\large \bf  Supplementary Information}
\end{flushleft}

\section{ALMA Observations}

%I think this looks good.  I'd add the quantum numbers for the other
%transitions as well.  Perhaps we could add '
%ą10%.'  That may be too many words but it informs where Titan's flux came from.

\subsection{Instrumental setup and data reduction}

\paragraph{ALMA setup.}
ALMA Band 7 observations of HD142527 were carried out in the night of
June 2 2012.  The precipitable water vapor in the atmosphere was
stable between 1.4 and 1.8 mm, with clear sky conditions. The ALMA
correlator was configured in the Frequency Division Mode (FDM) to
provide 468.750 MHz bandwidth in four different spectral windows at
122.07 kHz resolution (~0.1 km/s) per channel. Each spectral window
was positioned in order to target the CO(3-2) transition at 345.7959
GHz, HCO+ as well as CS(7-6) and HCN(4-3). The measured system
temperatures ranged from 207 to 285 K in the different spectral
windows. The number of 12~m antennas available at the time of the
observation was 19, although two antennas reported very large system
temperatures (DA41 and DV12) and were flagged during data
reduction. Excluding calibration overheads, a total time on source of
52 minutes was spent yielding an RMS of 15 mJy in 0.1~km~s$^{-1}$
channels. The primary flux calibrator was Titan, which provided a mean
transferred flux of 14.2~Jy for 3c279, the bandpass calibrator, and
0.55~Jy for J1604-446, the phase calibrator. Amplitude calibration
used the CASA Butler-JPL-Horizons 2010 model for Titan, which gives an
estimated systematic flux uncertainty of $\sim$10\%. All the line data
were processed with continuum subtraction in the visibility domain.

\paragraph{Image synthesis.}
Image synthesis was performed using two different techniques,
depending on the application. For a traditional way to present the
visibility dataset we use Cotton-Schwab {\tt CLEAN} in the {\tt CASA}
package. This technique represents the consensus in image
synthesis. We use Briggs weighting with robustness parameter of zero.
For deconvolved models we use a non-parametric least-squares modeling
technique\cite{2006ApJ...639..951C} with a regularizing entropy term
(i.e. as in the family of maximum entropy methods, MEM here and
elsewhere). MEM model images are restored by convolving with the clean
beam and by adding the residuals calculated using the {\tt difmap}
package\cite{Difmap}. For the residuals we use weights comparable to
our choice in {\tt CASA}, a mixture of natural and uniform weights. A
detailed example of this MEM algorithm is shown in the HCO$^+$ channel
maps, Fig.~\ref{fig:channels_HCOplus_MEM}.

%Fig.~\ref{fig:cont}.

%akin to Briggs weighting with robustness parameter of zero.

\begin{figure}
%\captionsetup{labelformat=bloub}
\begin{center}
\includegraphics[width=0.5\textwidth,height=!]{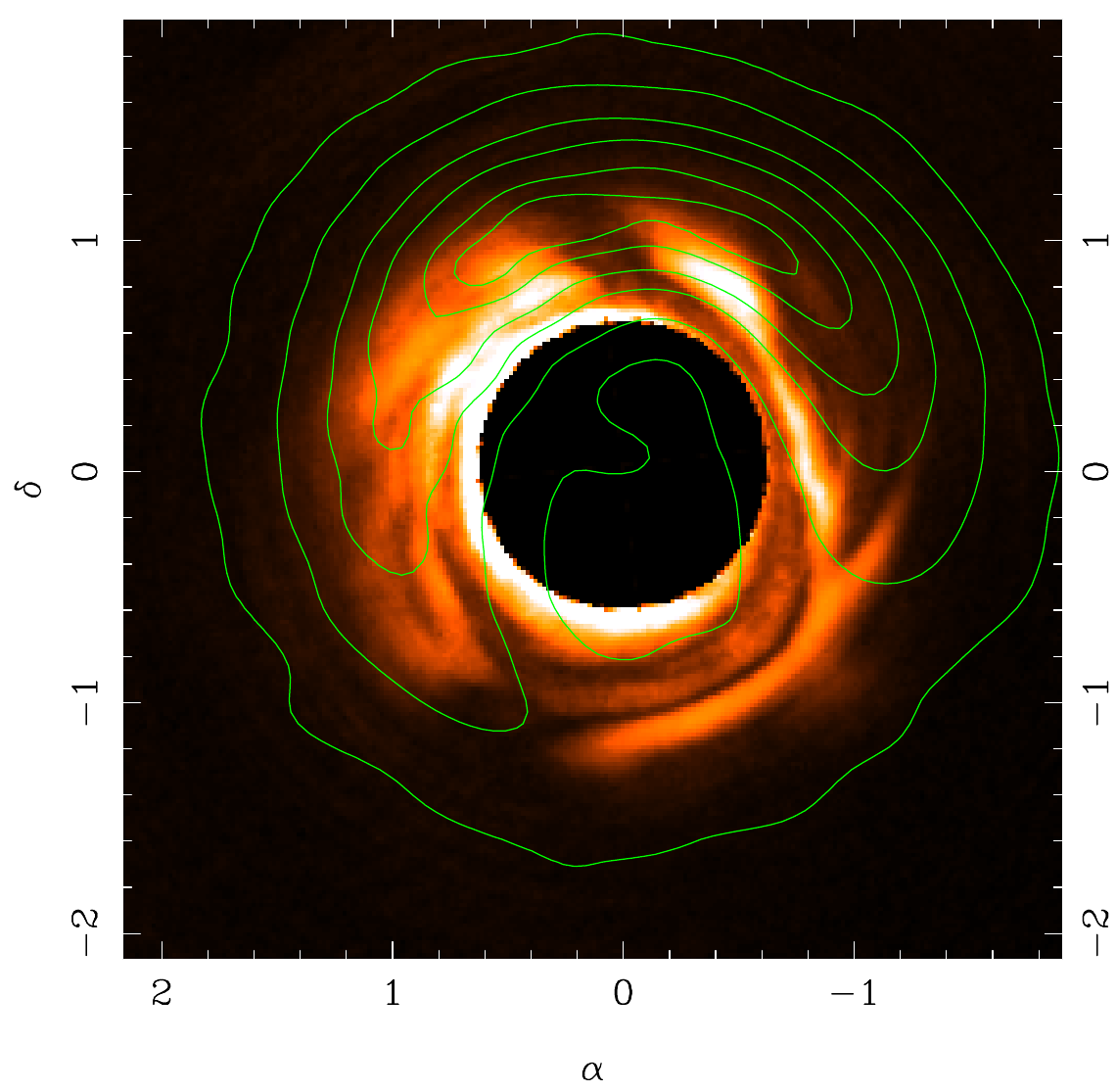}
\end{center}
\caption{{\bf Match between the radio continuum and near-IR disk}.
  Axis labels follow from Fig.~\ref{fig:cycle0}. We show 345~GHz
  continuum, from Fig.~\ref{fig:cycle0}a, overlaid on the 2~$\mu$m
  image from Fig.~\ref{fig:cycle0}c. Contour levels for the continuum
  are as in Fig.~\ref{fig:cycle0}a. 
 \label{fig:registration}}
\end{figure}

\paragraph{Registration of ALMA images.}   
A $\sim$0.1~arcsec astrometric uncertainty could affect the ALMA
data. However, we checked the astrometry by confirming that the
centroid of the Keplerian velocity field (seen in the RGB image for
CO(3-2) in Fig.~\ref{fig:cycle0}) lies indeed at the position of
HD~142527 in J2000 at the June 2012 epoch (J2000 15:56:41.878
-42:19:23.568). The near-IR scattered light
images\cite{Casassus2012ApJ...754L..31C}, which are centred on the
stellar position, are matched by the inner boundary of the sub-mm
thermal continuum, as expected. This match is illustrated in
Fig.~\ref{fig:registration}. Overall, we estimate that any astrometric
error is less than $\sim$0.05~arcsec.

\subsection{Self-calibration of continuum maps.}

A continuum image visibility dataset was constructed from the nominal
calibration of the source data, covering all ALMA spectral windows but
excluding line channels. The restored continuum image peaked at
0.38~Jy~beam$^{-1}$, with a noise level (including artifacts from
negative sidelobes) of 6~10$^{-3}$~Jy~beam$^{-1}$, giving a dynamic
range of 63.  The image was clearly limited by phase errors.  We
therefore applied self-calibration.  Using the nominal calibration
image as a model of the source, we used the self-calibration algorithm
to determine improved antenna-based phases that were consistent with
the image.  The solution interval chosen was 5 min, long enough to
give over 30:1 signal to noise (2 deg accuracy) per antenna solution.
The measured residual phases varied smoothly over the experiment, with
a typical value of 5 to 10 deg. After applying these small phase
corrections, the rms noise on the new image was 20\% of the original
rms, and the weak southern part of the outer disk became clearly
visible.  With this improved model, the next self-calibration
iteration also included the antenna gain variations which were
typically five percent or less.  This second round of self-calibration
decreased the rms a further 50\%.  The final image has a peak of
0.38~Jy~beam$^{-1}$ and an rms of $5.2~10^{-4}$~Jy~beam$^{-1}$, so of
the order of the theoretical noise level, which is 10\% of the
original rms. Since the dynamic range of the nominal calibration was
63, even the stongest line emission in 1~km~s$^{-1}$ channels, after
subtraction of the continuum emission, was receiver noise limited
rather than dynamic range limited.  Hence, self-calibration correction
to the line channels were not needed.

%Continuum visibilities at 345~GHz were extracted from the whole sets
%of ALMA spectral windows (excluding edges and the CO line).
%Fig.~\ref{fig:cycle0}a and Fig.~\ref{fig:contoverlay} show two
%different reconstructions of the same visibility dataset.  In
%particular the deconvolved model is better suited for comparison with
%the near-IR images. The most striking feature of the continuum is its
%horseshoe shape.  Differential stellar heating, of its own, cannot
%account for the observed north-south contrast of $>50$ at
%3~$\sigma$. The continuous HCO$^+$ ring supports that the ring has a
%clear view of the star, so that there is little UV opacity. In this
%case radiative thermal equilibrium predicts a $T\propto 1/\sqrt{r}$
%radius-temperature relationship. The ratio in radius to the inner edge
%of the disk, as traced by the near-IR image, is at most 1.4 in the
%north-south direction. Since thermal submm specific intensities are in
%the Rayleigh-Jeans regime and so proportional to the dust temperature,
%for differential stellar heating we would expect a meek modulation in
%continuum intensities of not more than $\sqrt{1.4}$.
%  

% \textcolor{red}{More on the continuum, comparison of CLEAN and
%  MEM-restored vs MEM-model to show that the MEM model really
%  deconvolves the elongation of the beam}

\subsection{Presentation of the ALMA data} \label{sec:alma}

\paragraph{Channel maps.}
The HCO$^+$ channel maps in Fig.~\ref{fig:channelsHCOplus} illustrate
the off-center high-velocity flows, and the gap-crossing filaments at a
velocity close to that of the star (the systemic velocity). These
filaments are seen more clearly in the deconvolved channel maps shown in
Fig.~\ref{fig:channels_HCOplus_MEM}. The CO(3-2) channel maps shown in
Fig.~\ref{fig:channelsCO} also highlight the blue-shifted part of the
high-velocity flows seen in HCO$^+$.

\paragraph{HCO$^+$ peak and intensity maps.}
We use the same set of CLEANed HCO$^+$ images from 5-channel averages
as in the channel maps to summarize the basic features of the HCO$^+$
emission through moment maps.  Fig.~\ref{fig:whcomax}a exposes in more
detail the total intensity map shown in Fig.~\ref{fig:cycle0}. The
peak intensity map in Fig~\ref{fig:whcomax}c illustrates that the
outer disk is a whole ring in HCO$^+$.

\begin{figure}
\begin{center}
\includegraphics[width=0.9\textwidth,height=!]{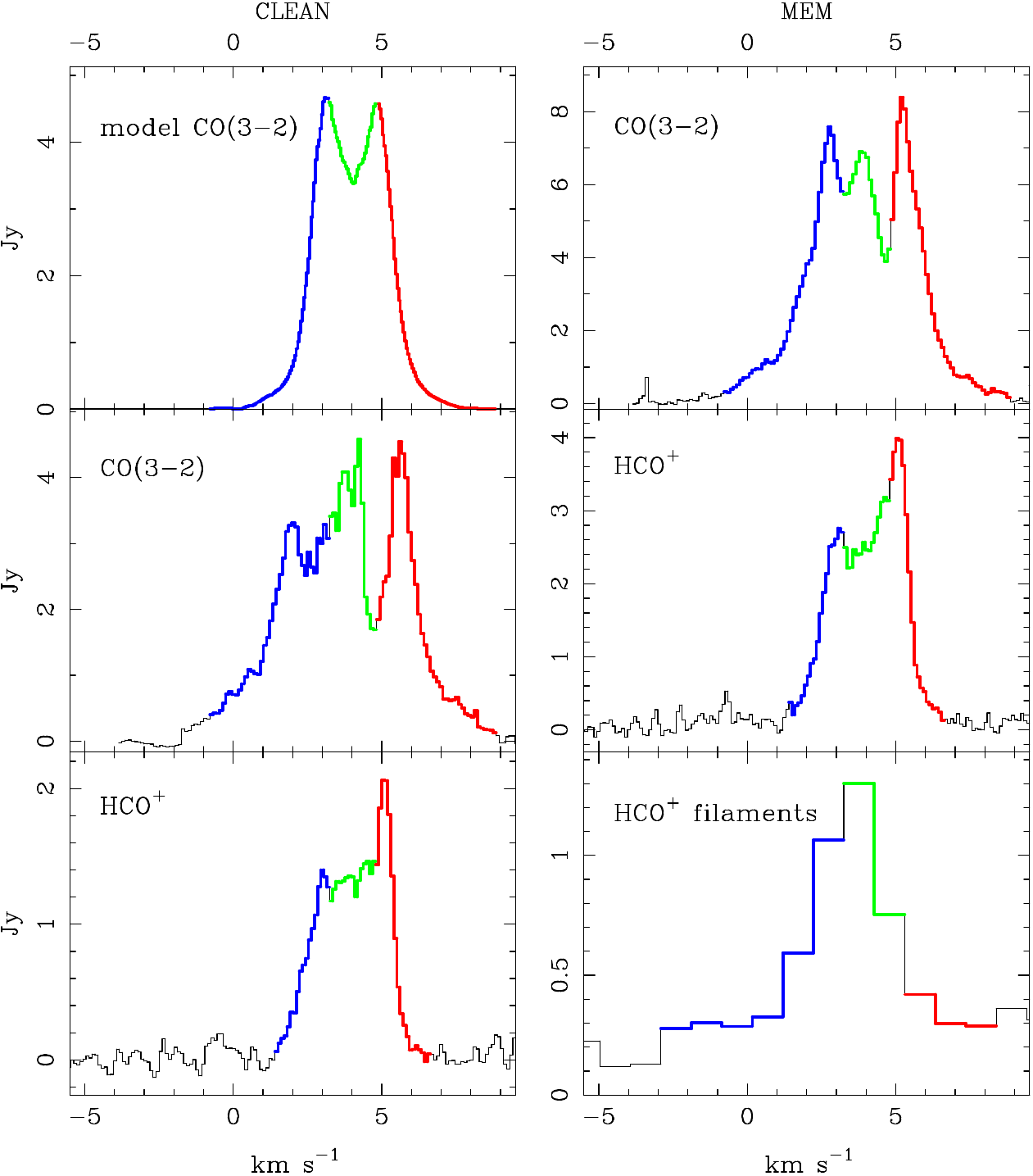}
\end{center}
\caption{{\bf Spectra extracted from the RGB images shown in
    Fig.~\ref{fig:cycle0}}. These spectra are meant to inform the
  definition of the RGB velocity codes, and are extracted over
  apertures defined by the field of view of the comparison RGB
  images. These spectra do not represent accurate measures of the
  total integrated spectrum. Comparison with previous
  work\cite{2011ApJ...734...98O} at coarser resolutions shows that these
  extractions are modulated by flux loss and CLEAN artifacts. Left
  column shows CLEAN reconstructions, while right column shows MEM
  reconstructions.  Labels indicate flux density in $y-$axis, in Jy,
  and $v_\mathrm{LSR}$ in $x-$axis, in km~s$^{-1}$. Notice the
  irregular blue peak in CO(3-2), which is modulated by CLEAN
  artefacts (extended negatives to the north). Our MEM recovers more
  flux but is noisier in individual channels of this dataset. Note
  also that the comparison model CO(3-2) spectrum is sensitive on the
  adopted temperature profile, so is shown here only as a reference
  Keplerian rotation model.  The spectrum labels correspond to the
  following figures: `model CO(3-2)', Fig.~\ref{fig:lime}; `CO(3-2)',
  Fig.~\ref{fig:cycle0}\,b; `HCO$^+$', Fig.~\ref{fig:cycle0}\,d;
  `filaments', inset to Fig.~\ref{fig:cycle0}\,d. 
 \label{fig:spectra}}
\end{figure}

\paragraph{Line spectra.}
The first column in Fig.~\ref{fig:spectra} shows spectra extracted from
CLEAN reconstructions, using the same aperture as show on
Fig.~\ref{fig:cycle0}. The right column shows the corresponding MEM
spectra. The spectrum labelled `streamer', in bottom right, corresponds
to the inset of Fig.~\ref{fig:cycle0}\,d.

\section{Comparison disk models} \label{sec:lime}

To illustrate the expected kinematics in Keplerian rotation we have
calculated a disk model using the LIME\cite{2010A&A...523A..25B}
package. This model is inclined at 20~deg, with a PA of
$-20$~deg east of north, and is placed at a distance of 140~pc
with a 2.7~M$_\odot$ central star. The distance of 140~pc is supported
by a probable association to Sco~OB2.

For CO we assumed an abundance of $10^{-4}$ relative to H$_2$ and the following
number density distribution of H$_2$ molecules:
\[
n_\mathrm{H2}(r,z)   =   1.5~10^{14}~\left( \frac{r}{100~\mathrm{AU}} \right)^{-1.5}~\exp\left[  - \frac{1}{2} \left( \frac{z}{0.1~r} \right) ^2  \right] ~ \mathrm{m}^{-3}, 
%end{eqnarray}
\]
from 140 to 300~AU, and 
\[
n_\mathrm{H2}(r,z)   =   \zeta ~ 1.5~10^{14}~\exp\left[  - \frac{1}{2} \left( \frac{z}{0.1~r} \right) ^2  \right] ~ \mathrm{m}^{-3}, 
%end{eqnarray}
\]
from 10 to 140~AU, and zero elsewhere. $\zeta = 1/10$ is a fiducial
well-depth. The total mass in this disk model is 0.1~M$_\odot$.  The
temperature profile is
$T(r)=50~\left[r/(50~\mathrm{AU}\right]^{-1/2}~K$, as inspired from
the observed CO(3-2) peak with a radiation temperature of 50~K.

For HCO$^+$ we assumed a relative abundance of $10^{-7}$, a mass
distribution as for CO but without material inside the gap, and the
same temperature profile.

\begin{figure}
\begin{center}
\includegraphics[width=0.5\textwidth,height=!]{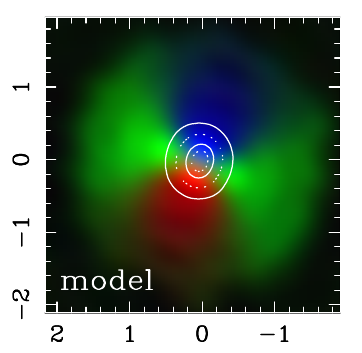}
\end{center}
\caption{{\bf Prediction for CO(3-2) from a fiducial disk model for
    comparison with the velocity field coded in RGB in
    Fig.~\ref{fig:cycle0}}.  This prediction has been filtered by the
  ALMA $uv$-coverage and reconstructed with CLEAN, in the same way as
  the observations. Axis labels and color codes follow from
  Fig.~\ref{fig:cycle0}b.
\label{fig:lime}}
\end{figure}

We filtered the LIME predictions to match the $uv$-coverage of the
data, and we reconstructed the resulting visibility in the same way as
the observations. Resuts for CO(3-2) are shown in Fig.~\ref{fig:lime}
and Fig.~\ref{fig:channelsCO}.  In the case of HCO$^+$ the line is
weaker so we subtracted the continuum. We also added a central
elongated source to assess the effects of beam elongation, see
Fig.~\ref{fig:channelsHCOplus}.  We conclude that beam elongation
effects cannot mimick the HCO$^+$ filaments.

%Under these assumptions the peak CO radiation temperature corresponds
%to a physical temperature of 50~K (see Sec.~\ref{sec:conditions}). We
%proceeded to calculate $^{12}$CO(3-2) model predictions for a
%parametrised model disk that reproduces the observed mass of
%0.1~M$_\odot$, and with two different gap depth values
%
%two different depth values.
%

\section{Physical conditions} \label{sec:conditions}

% Comparing the HCO$^+$ and CO signals we infer limits on the column of
% hydrogen-nuclei of $ 5~10^{22} > N_H/$cm$^{-2}> 10^{20}$ and a
% temperature of 50~K (see Sec.~\ref{sec:conditions} in SI).  The
% continuum levels on the Easter

\paragraph{Conditions along the filaments}
An estimate of physical conditions in the filaments can be obtained by
assuming that the CO(3-2) emission is optically thick. If so the peak
CO(3-2) specific intensity of 0.9~Jy~beam$^{-1}$ in a
$0.55\times0.33$~arcsec$^{2}$ beam corresponds to a kinetic
temperature of 50~K.  The peak HCO$^+$ specific intensity is
0.1~Jy~beam$^{-1}$ or $\sim$5.7~K in radiation temperature.  This
implies HCO$^+$ columns of $N_\mathrm{HCO+} \sim
5~10^{14}$~cm$^{-2}$. Densities in excess of $n_{\mathrm{H}2} \sim 10
^{5}$~cm$^{-3}$ are required to excite HCO$^+$(4-3) and if the
streamers are as deep as they are wide ($\sim 5~10^{14}$~cm
corresponding to 0.25~arcsec), we expect H-nucleus columns $ N_H >
10^{20}$cm$^{-2}$. We thus have an upper bound on the HCO$^+$
abundance of $N_\mathrm{HCO+} / N_H < 5~10^{-6}$. A lower bound is
placed by the observed (cosmic-ray induced) ionization fraction in
cold cores\cite{Wootten1979ApJ...234..876W} of $\sim 10^{-8}$. In
summary, from HCO$^+$ we have $ 5~10^{22} \gtrsim N_H/$cm$^{-2}
\gtrsim 10^{20}$. The observed continuum under the filaments ranges
from 0.4~mJy~beam$^{-1}$ to 5.6~mJy~beam$^{-1}$ before beeing confused
with the wall, and the corresponding columns are $ 1.3~10^{23} \gtrsim
N_H/$cm$^{-2} \gtrsim 8.5~10^{21}$. Extinction values can be estimated
using a standard formula\cite{Draine2003ARA&A..41..241D},
$N_H/$cm$^{-2} = 1.87~10^{21} A_V$.

\begin{figure}
\begin{center}
\includegraphics[width=0.7\textwidth,height=!]{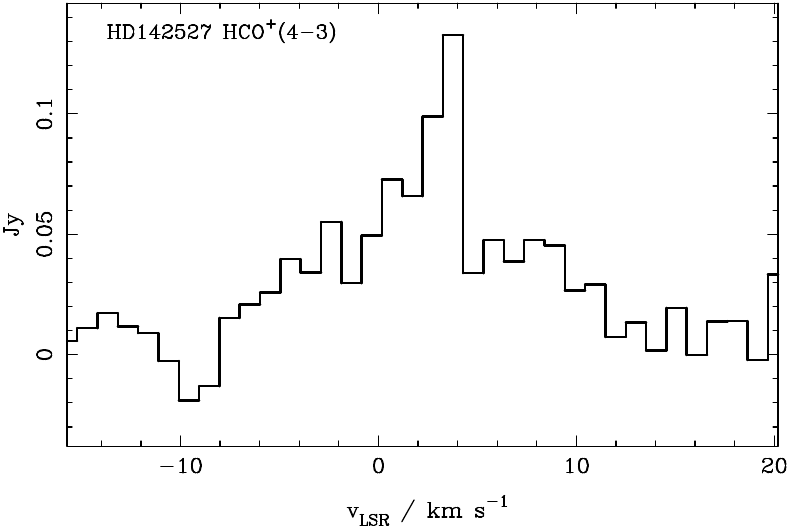}
\end{center}
\caption{{\bf HCO$^+$ spectrum extracted from the central intensity
    peak}. The aperture size is $0.5\times0.6$~arcsec. 
 \label{fig:centralspec} }
\end{figure}

\paragraph{Physicality of central outflows} 
On first impression, star formation experts would interpret the
non-Keplerian HCO$^+$ as a stellar outflow. The orientation of the
eastern filament is indeed roughly orthogonal to the disk PA, and
outflows are expected in the context of stellar accretion. However,
planet-induced accretion may not require outflows to dissipate angular
momentum (whose bulk has been dissipated earlier by powerful stellar
outflows during the general protostellar accretion).  Outflows have
optical/IR counterparts, but no extended emission is seen in our
Br$\gamma$ imaging from Gemini (SI). In general outflows are not seen
in transition disks\cite{Sacco2012ApJ...747..142S} (class~II young
stars with disks and an inner cavity).  Here we consider the
plausibility of a ballistic outflow interpretation for the
filaments. Given an inclination of 20~deg, the most conspicuous part
of the filaments, i.e. the central HCO$^+$ unresolved intensity peak,
would lie $<80~$AU above the star (from the maximum possible angular
distance to the star of the blue-shifted and off-center component), at
velocities $<7.2$~km~s$^{-1}$ relative to the star. The bulk of the
material flows at lower velocities, in fact the peak is at systemic
velocities (Fig.~\ref{fig:centralspec}). Yet the line-of-sight
velocity spread of HCO$^+$ emission from the outer disk, of
4~km~s$^{-1}$ at 1.0~arcsec, corresponds to rotation velocities of
5.9~km~s$^{-1}$ at 140~AU in the plane of the disk. The corresponding
escape velocity at 80~AU is 8.6~km~s$^{-1}$. So the bulk of this
outflow cannot be ballistic (in particular the systemic-velocity
filament, which extend from the star and into the outer disk).
Therefore an ambient medium would be required to decelerate this
hypothetical outflow. In this scheme a jet would entrain ambient
molecular gas and produce the observe HCO$^+$ flows. However, no
optical/IR jet is seen.  A Br$\gamma$ line is conspicuous in the
stellar spectrum, but it is unresolved (see Fig.~\ref{fig:NICI}), so
contained well within the inner 5~AU. Additionally, this molecular
medium would need to be pressure-supported, as in a class~I envelope,
and extend vertically above the disk to $\sim$100~AU. Yet the observed
SED\cite{2011A&A...528A..91V} shows very little extinction ($A_V
\lesssim 0.6$), consistent with interstellar values rather than
intranebular extinction.

\paragraph{Physicality of an orthogonal inner disk} 
The high-velocity central flows could be thought to stem from
Keplerian rotation orthogonal to the plane of the outer disk, so with
an inclination of $\sim$90~deg (high inclinations are required to
account for aspect ratio). However, given the outer-disk Keplerian
velocities, an inner disk extending out to 0.1~arcsec but orthogonal
to the plane of the outer disk should show a double-peaked spectrum
with line of sight velocities $\sim 26~$km~s$^{-1}$, while the
observed peak is at 7.2~km~s$^{-1}$. In addition to these dynamical
arguments, an orthogonal inner disk is also inconsistent with the
observed CO~4.67~$\mu$m emission. This fundamental ro-vibrational
transition is seen to originate from an inner disk that is aligned
with the outer disk\cite{Pontoppidan2011ApJ...733...84P}.

\paragraph{Estimates of mass inflow rates}  
Under the hypothesis that the observed filaments are accretion
streams, we can estimate a mass inflow rate onto the inner disk. A
lower limit comes from the HCO$^+$(4-3) observation. This line stems
from gas close to the critical density, so close to $n_\mathrm{H2}
\sim 10^6$~cm$^{-3}$. The average density along the filaments is
probably higher. With an inflow velocity of 5~km~s$^{-1}$, and a
filament section of $\sim 0.25\times0.25$~arcsec$^2$, we obtain a mass
inflow rate of $dM_i/dt > 7~10^{-9}~$M$_\odot$~yr$^{-1}$.  The mass
seen in the dust continuum provides another limit. The fraction of
flux in the filaments is $1/1000$, so the filaments carry $\sim
10^{-4}$M$_\odot$ for a uniform and standard dust to gas ratio. For a
characteristic radial velocity $< 1$km~s$^{-1}$ (we do not detect the
radial velocity of the filaments - only their roots are infalling),
and for a liner size of 100~AU, the dynamical age of the filaments is
$> 474$~yr. The corresponding mass infall rate is $dM_i/dt <
2~10^{-7}~$M$_\odot$~yr$^{-1}$. We see that the observed stellar
accretion rate\cite{2006A&A...459..837G}, of $7 \times
10^{-8}$~M$_{\odot}$~yr~$^{-1}$, is bracketed within one order of
magnitude by the above limits on the filament mass inflow rate.

\paragraph{North-south asymmetry in the continuum.} 
The most striking feature of the continuum is its horseshoe shape.
Differential stellar heating, on its own, cannot account for the
observed north-south contrast of $28\pm0.5$. The continuous HCO$^+$
ring supports that the ring has a clear view of the star, so that
there is little UV opacity. In this case radiative thermal equilibrium
predicts a $T\propto 1/\sqrt{r}$ radius-temperature relationship. The
ratio in radius to the inner edge of the disk, as traced by the
near-IR image, is at most 1.4 in the north-south direction. Since
thermal submm specific intensities are in the Rayleigh-Jeans regime
and so proportional to the dust temperature, for differential stellar
heating we would expect a meek modulation in continuum intensities of
not more than $\sqrt{1.4}$.

\section{Near-IR ADI processing and search for protoplanets} \label{sec:ADI}

%Channel1 (Red): CH4-K5%L G0748 (2.241 um with 107 nm bandwidth, CH4L hereafter)
%Channel2 (Blue): CH4-K5%S G0746 (2.080 um with 105 nm bandwidth, CH4S hereafter)

%HD~142527 was imaged by the the Near-Infrared Coronagraphic Imager
%(NICI) at the Gemini South telescope with the CH4-K5\%L\_G0748 and
%CH4-K5\%S\_G0746 filters,

HD 142527 was imaged by the Near-Infrared Coronagraphic Imager (NICI)
at the Gemini South telescope in the two filters, CH4-K5\%L\_G0748
(2.241 $\mu$m with 107 nm bandwidth, CH4L hereafter) and
CH4-K5\%S\_G0746 (2.080 $\mu$m with 105 nm bandwidth, CH4S hereafter),
using a $0.22$~arcsec semi-transparent coronagraphic mask
(with 95\% Lyot stop) and in pupil tracking mode (the rotator is
disabled to let the field rotate and to stabilize the pupil). The
total PA rotation was $\sim$40~deg. We also obtained unsaturated
Br$\gamma$ data, with a $\sim$20~deg rotation.  Seeing and
transparency conditions were good.

We reduced both CH4 channels using a set of pipelines run
independently and in parallel. Each pipeline was sequentially optimized to
retrieve either the disk extended emission and/or point sources.

First of all, it must be noted that face-on disks are particularly
difficult to recover using ADI
techniques\cite{Milli2012arXiv1207.5909M}, which is the case of
HD~142527. Furthermore the complex disk structure makes ADI processing
even more prone to artifacts.  In order to obtain a confident view of
the disk, we decided to perform statistical PSF averaging instead of
PSF subtraction (Fig.~\ref{fig:cycle0}\,c), using the Geneva PADIP
pipeline.  Apart from PSF subtraction we kept all the other reduction
steps, namely flat fielding, bad pixel removal, high-pass filtering,
strehl based frame selection, Fourier based recentring and derotation.
This way, by skipping the PSF subtraction step we get rid of most
artifacts common to all ADI algorithms due to disk self-subtraction,
but at the cost of a lower contrast and an increased inner working
angle at $0.7$~arcsec. The resulting image is a median of the
best frames, corrected for field rotation.

\begin{figure}
\begin{center}
\includegraphics[width=0.5\textwidth,height=!]{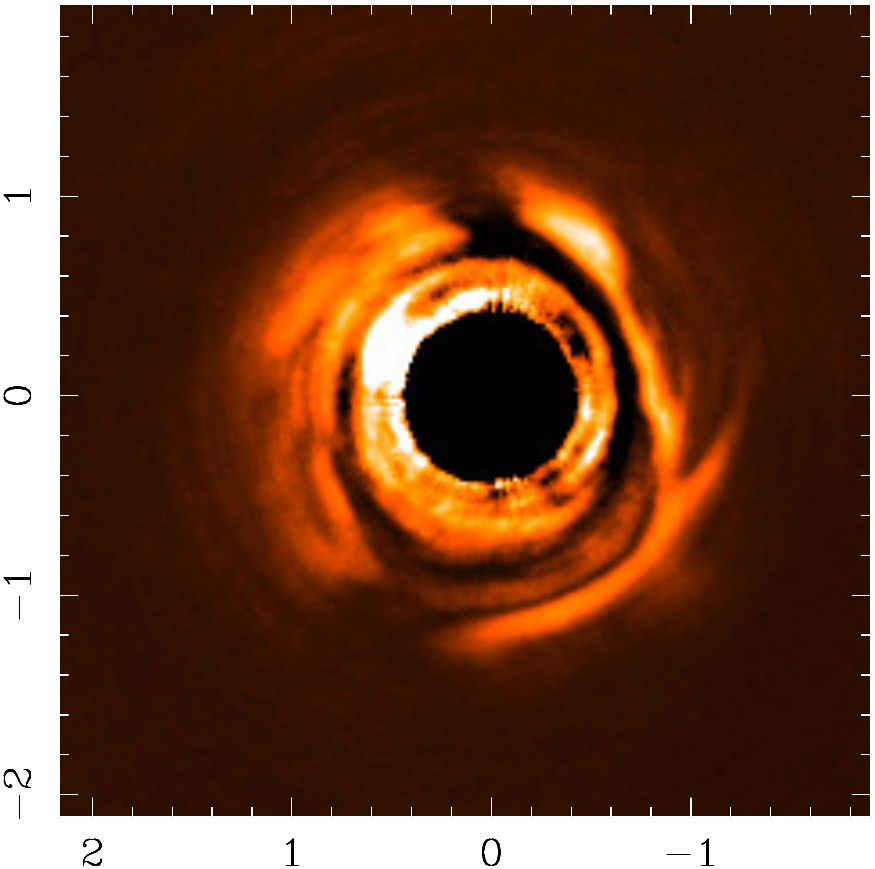} % fig_NICIdisk.pl
\end{center}
\caption{{\bf Gemini NICI image of the disk with approximate
    PSF-subtraction}. This image shows the CH4S filter, with PSF
  averaging and derotation.  It shows the least biased disk image, as
  Fig.~\ref{fig:cycle0}\,c, but with approximate PSF subtraction using
  an azimuthally-varying Moffat profile fit to the PSF halo.  The
  spiral structure seen inside the cavity (immediately surrounding the
  intensity mask) is probably the result of strong static speckles in
  these ADI data.
\label{fig:Moffat}}
\end{figure}

In order to extract further details from the disk image with
statistical PSF averaging we subtracted a Moffat profile fit to the
stellar PSF over distinct regions in azimuth. The result is shown on
Fig.~\ref{fig:Moffat}. The final result of this conservative reduction
is the most detailed view of the HD~142527 complex disk structure ever
obtained (the near-IR images have finer resolution than Cycle~0 ALMA
data). We caution that the arc-like structures inside
$0.7$~arcsec are probably artifacts. As they rotate in
parallactic angle, the strongest static speckles can bias the median
disk image into such arc-like features.

\begin{figure}
\begin{center}
\includegraphics[width=0.9\textwidth,height=!]{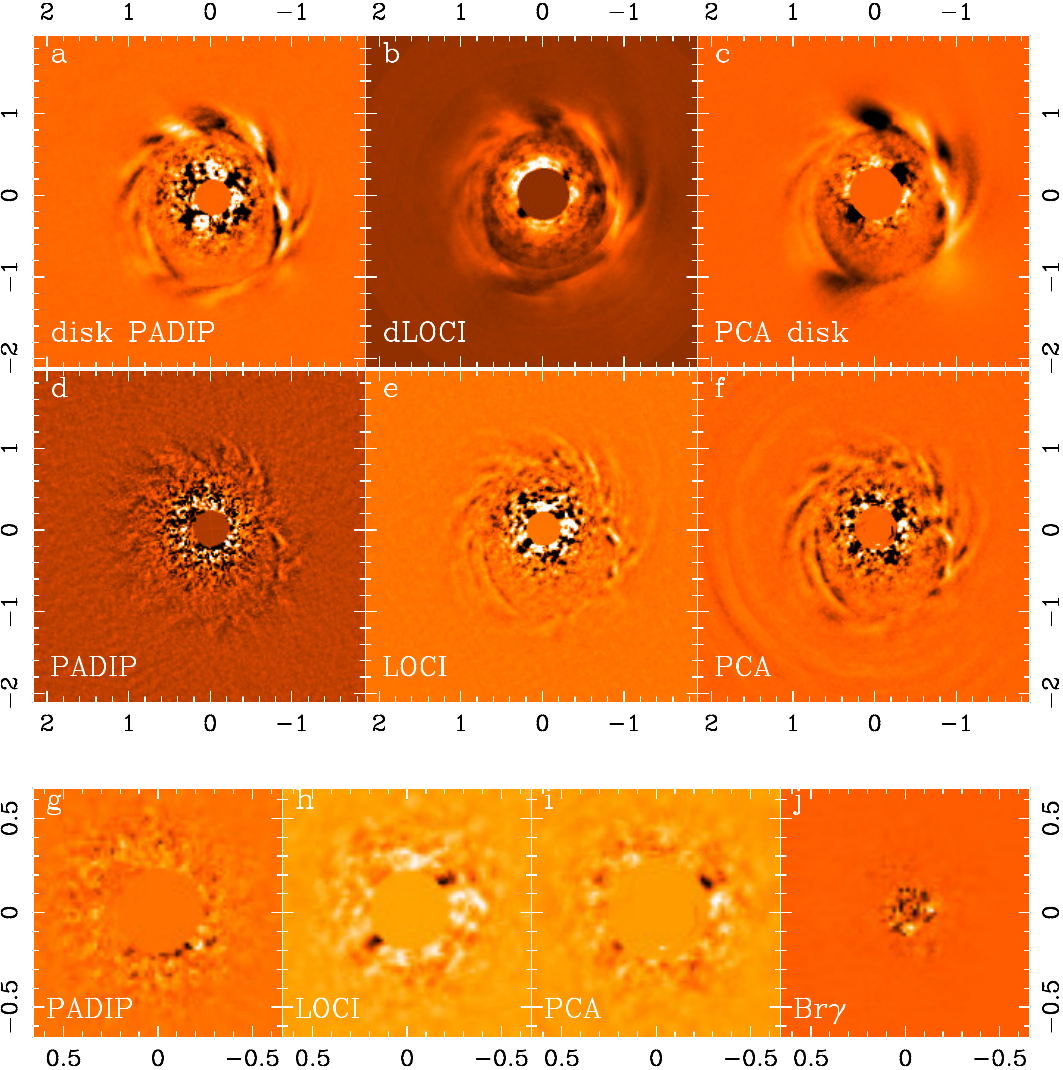} % fig_ADI.pl
\end{center}
\caption{{\bf Gemini NICI images after various PSF-subtraction
    methods}. All panels show the CH4S filter (an analogous set of
  images is available in CH4L), except for j). {\bf a:} PADIP pipeline
  image of the disk. {\bf b:} Damped-LOCI image of the disk. {\bf c:}
  PCA image of the disk.{\bf d:} Aggressive PADIP image for faint
  companion search. {\bf e:} Aggressive LOCI image.  {\bf f:}
  Aggressive PCA image. {\bf g:} Aggressive PADIP zoom. {\bf h:}
  Aggressive LOCI zoom. {\bf i:} Aggressive PCA zoom. {\bf j}:
  Br$\gamma$ RDI limits close-in stellar companions - this last image
  was reduced with PCA, its IWA is only 50~mas, as opposed to 220~mas
  in CH4S.

\label{fig:NICI}}
\end{figure}

To optimize the search for point sources around the star within the
disk gap we then ran pipelines in conventional ADI mode with
sophisticated PSF subtraction algorithms.  PSF subtraction had to be
used in order to have access to the inner region of the disk inside
$0.7$~arcsec. This was achieved by running three different
pipelines.  The PSF reconstruction was implemented by a set of
methods based respectively on the well-known
ADI\cite{Marois2006ApJ...641..556M} PSF subtraction technique (PADIP),
the locally optimized combination of images
(LOCI\cite{Lafreniere2007}), and the new principal component analysis
(PCA\cite{2012ApJ...755L..28S}). The task is particulary complex due
to the disk structure in the outer regions, and to the presence of
strong persistent speckles close to the star  as can be seen on
Fig.~\ref{fig:NICI}\,b--f.  It should be noted that artifacts caused
by the ADI observing strategy do not propagate radially when reduced
with all methods except for LOCI, which means that the outer disk
structures will not induce artifacts in the inner cavity for all
methods except LOCI.

PADIP (Fig.~\ref{fig:NICI}\,d,g) performs parallelised frame
operations in the Fourier space, with high-pass pre-filtering,
subtracting optimized PSFs for each single frame, and setting priority
on flux conservation. It was initially conceived to find point
sources, but a new mode optimized for disk reductions, which reduces
the ADI induced artifacts by using smeared PSF references, has been
implemented for these observations.

The second method we used, LOCI (Fig.~\ref{fig:NICI}\,e, h), finds the
optimal linear combination of reference frames to minimize the noise
in a given zone of the target image. The process is repeated until the
area of interest in the target image is completely reduced. LOCI has a
known tendency to generate artifacts in extended sources such as
circumstellar disks. However, this defect of the generic LOCI
algorithm is brought under control in a modified version.  The d-LOCI
algorithm\cite{Pueyo2012,Mawet2012arXiv1207.6017M} (Fig.~\ref{fig:NICI}\,b)
incorporates a fine tuning of the geometrical parameters in LOCI (such
as the size of the optimization zone, the number of reference frames
used in the correlation matrix, as well as the introduction of a
damping parameter through a Lagrange multiplier) to balance flux
conservation with noise attenuation.

The third PSF subtraction method, based on PCA
(Fig.~\ref{fig:NICI}\,f,j)), proceeds as follows: assuming a library of
reference PSFs, a Karhunen-Lo\`eve transform of these references is
used to create an orthogonal basis of eigenimages, on which the
science target is projected to create the reference PSF. A PSF
constructed in this fashion minimizes the expected value of the
least-squares distance between the ensemble of reference images and
the random realization of the telescope response contained in the
science image.

%, one being at the expected position of the announced close stellar
%companion (see Sec~\ref{sec:biller} below)

Small inner working angles are difficult to obtain with ADI techniques
because the PA variation needed close in is too constraining. We also
obtained unsaturated Br$\gamma$ data sets with NICI. This data set,
although taken in pupil tracking mode (PA modulation of $\simeq 20$~deg), was also accompanied by a standard star, that was used to
construct a reference PSF using the PCA method for reference star
differential imaging (RDI). The corresponding image is shown in
Fig.~\ref{fig:NICI}\,k.

\begin{figure}
\begin{center}
\includegraphics[width=0.9\textwidth,height=!]{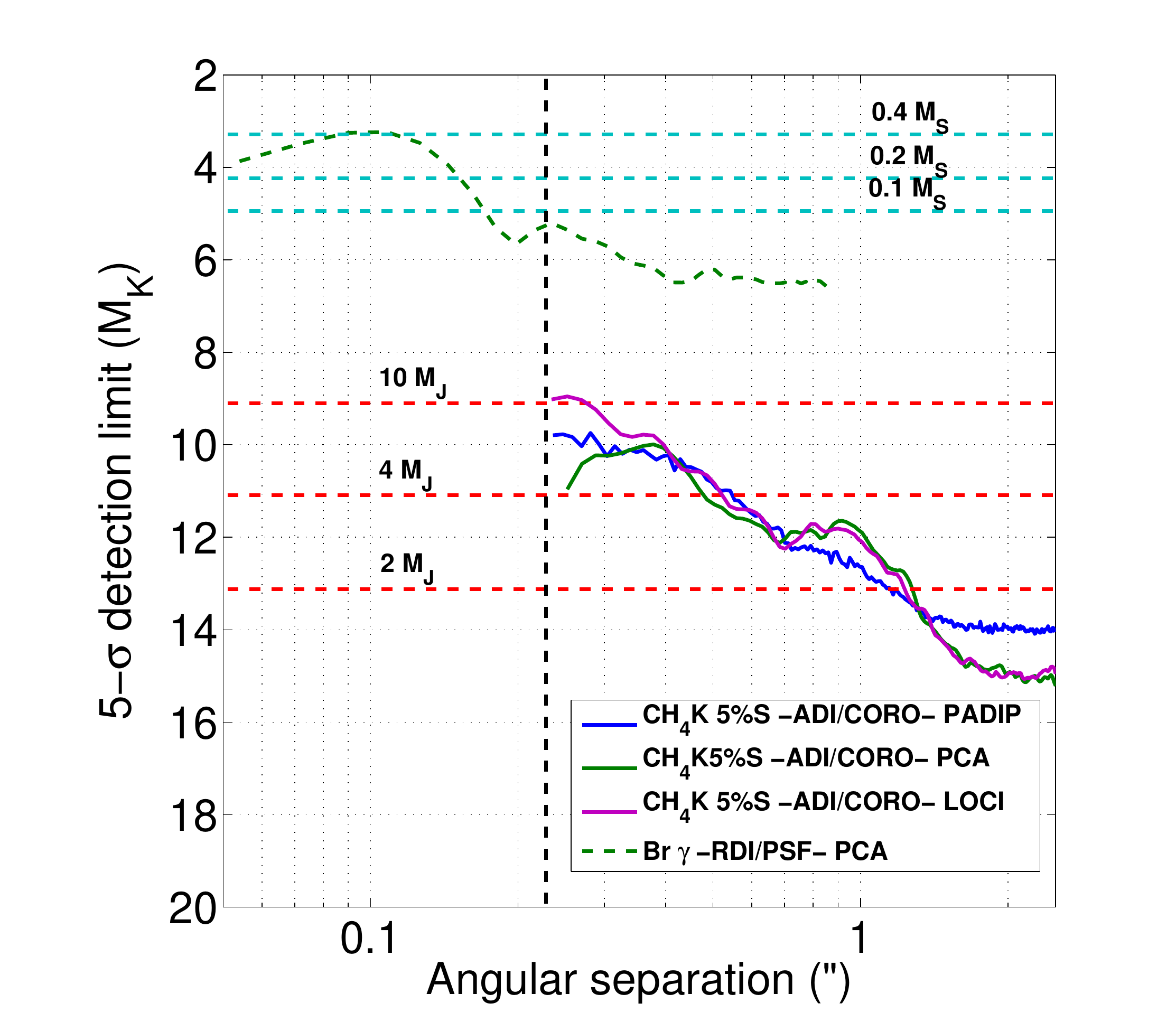}
\end{center}
\caption{{\bf Limits on companion mass from NICI CH$_4$ ADI, and NICI
    $Br\gamma$ reference star differential imaging (RDI)).} For NICI
  CH$_4$ ADI, we report contrast curves in three different techniques
  (PADIP, LOCI and PCA, see text for details). These limits are not
  corrected for extinction. As a comparison point we refer to the {\tt
    COND03}\cite{Baraffe2003} tracks at 2~Myr of age, which lead to
  the mass upper limits indicate on the plots. The vertical dashed
  black line indicates the effective inner working angle, set here by
  the extent ($0.22$~arcsec) of the semi-transparent Lyot coronagraph
  used in the NICI CH$_4$ ADI data set. The NICI $Br\gamma$ detection
  limits obtained with RDI are also indicated along with 2~Myr stellar
  tracks using the BCAH98
  model\cite{Baraffe1998A&A...337..403B}.  \label{fig:ADI}}

\end{figure}

No obvious point sources could be detected by either methods. Several
hot spots are nevertheless identified, but additional follow-up to
further characterize them is required.  At this point it is not clear
if these hot spots are related to the disk, bridges, or putative
companions currently forming within the disk. Based on these
reductions, we derived conservative upper limits, summarized in
Fig.~\ref{fig:ADI}, which also give new upper limit close stellar
companions within 0.1~arcsec (such as HD~142527B tentatively reported
from optical interferometry, see Sec~\ref{sec:biller} below).

%excluding the hot spots, 
%as summarized in Fig.~\ref{fig:ADI}.

%Despite our considerable data analysis efforts, the
%average quality of this set did not allow us to reach the detectivity
%level required to image the putative HD142527B (Fig.~\ref{fig:ADI}),
%even though current capabilities should not prevent us to reach this
%regime in the near future.
%

%SIMON: suggested caption for ADI panel figure:
%Panel showing the final images our set of pipeline produced for the NICI $CH_4$ ADI data. a: PSF averaging and derotation, showing the least biased disk image, but affected by strong PSF residuals. b: aggressive LOCI image, focusing on faint companion detection. c: damped-LOCI image, as first step towards the disk flux conservation vs PSF subtraction trade-off. d: PCA analysis, representing the best compromise between detection capabilities and flux conservation.

%NOTE: as we are in the supplemental material section, I took the freedom to add many details. The idea, which I hope you share is to show both the editor and the referees that we did our best to make the most of the data...

\section{Physical considerations on close stellar companions}  \label{sec:biller}

A non-zero closure phase from sparse-aperture-masking (SAM) near-IR
data has recently\cite{Biller2012} been interpreted in the context of
binary models, which are optimal for a 0.1--0.4~M$_\odot$ companion at
$\sim$13~AU (88~marcsec). Our RDI data limit such a companion to less
than 0.3~M$_\odot$.  The reality of this stellar companion is
debatable, as the binary model for the visibility data ignores the
inner disk, which accounts for the largest fraction of the near-IR
flux\cite{2011A&A...528A..91V}. We tested the binary model by
simulating SAM observations at 2~$\mu$m on a radiative transfer
prediction obtained with the MCFOST\cite{2006A&A...459..797P} package,
for an azimuthally symmetric model inner disk that is consistent with
the spectral energy distribution (SED). We find that even after
Fourier-filtering, the visibilities from the disk in the SAM $u,v$
coverage reach $\sim$2~Jy (for comparison values we refer the reader
to radiative transfer modeling of the observed
SED\cite{2011A&A...528A..91V}), while the flux density from a
hypothetical HD~142527B would be down at $\sim$0.065~Jy (in $K$, for a
magnitude difference of 4.8 and a total magnitude of 5.0). Thus
deviations from axial symmetry amounting to a mere 3.25\% of the total
inner disk flux could account for the phase closures that have been
interpreted in a binary model. The dust scattering phase function is
inherently asymmetric, so that radiative transfer effects alone can
reproduce the observed closure phases, even based on axysymmetric disk
models.  In any event, at such short separations this putative stellar
companion would lead to a very different disk morphology. The cavity
would be much smaller\cite{Casassus2012ApJ...754L..31C}, and be
entirely devoid of gas.

\begin{center}
{\bf ADDITIONAL REFERENCES}
\end{center}

\bibliography{/Users/simon/common/texinputs/merged.bib}
%\bibliography{/b/home/simon/common/texinputs/merged.bib}

\newpage

\newpage

%\begin{figure}
%\begin{center}
%\includegraphics[width=0.9\textwidth,height=!]{cont.pdf}
%\end{center}
%\caption{{\bf Deconvolved models of the continuum at 345~GHz
%    illustrate the removal of beam elongation effects}.  Axis labels
%  for images are in arcsec; North is up, East is to the left. The
%  origin of coordinates is centred on the star. The beam size is $
%  0.51 \times 0.33$~arcsec$^{-2}$ (as in Fig.~\ref{fig:cycle0}). In
%  both images overlaying contour levels are taken on the background
%  images, at the following fraction of peak intensity, 0.3, 0.5, 0.75,
%  and 0.9.  {\bf a:} restored map from the deconvolved mode shown in
%  b), in exponential grey scale, and with contour levels comparable to
%  Fig.~\ref{fig:cycle0}a, with a peak intensity of
%  0.334~Jy~beam$^{-1}$. The noise, as estimated from the
%  root-mean-square dispersion of the (primary-beam-corrected)
%  residuals is 1~$\sigma$=2.0~mJy~beam$^{-1}$.  {\bf b:} Corresponding
%  deconvolved image, notice that the width of the ring, i.e. the full
%  width at half maximum taken radially from the star, is fairly
%  constant. This contrasts with Fig.~\ref{fig:cycle0}a, where a
%  broadening of the ring is seen in the NE, corresponding to the
%  direction of the beam major axis. 
% \label{fig:cont}}
%\end{figure}

\begin{figure}
\begin{center}
\includegraphics[width=0.9\textwidth,height=!]{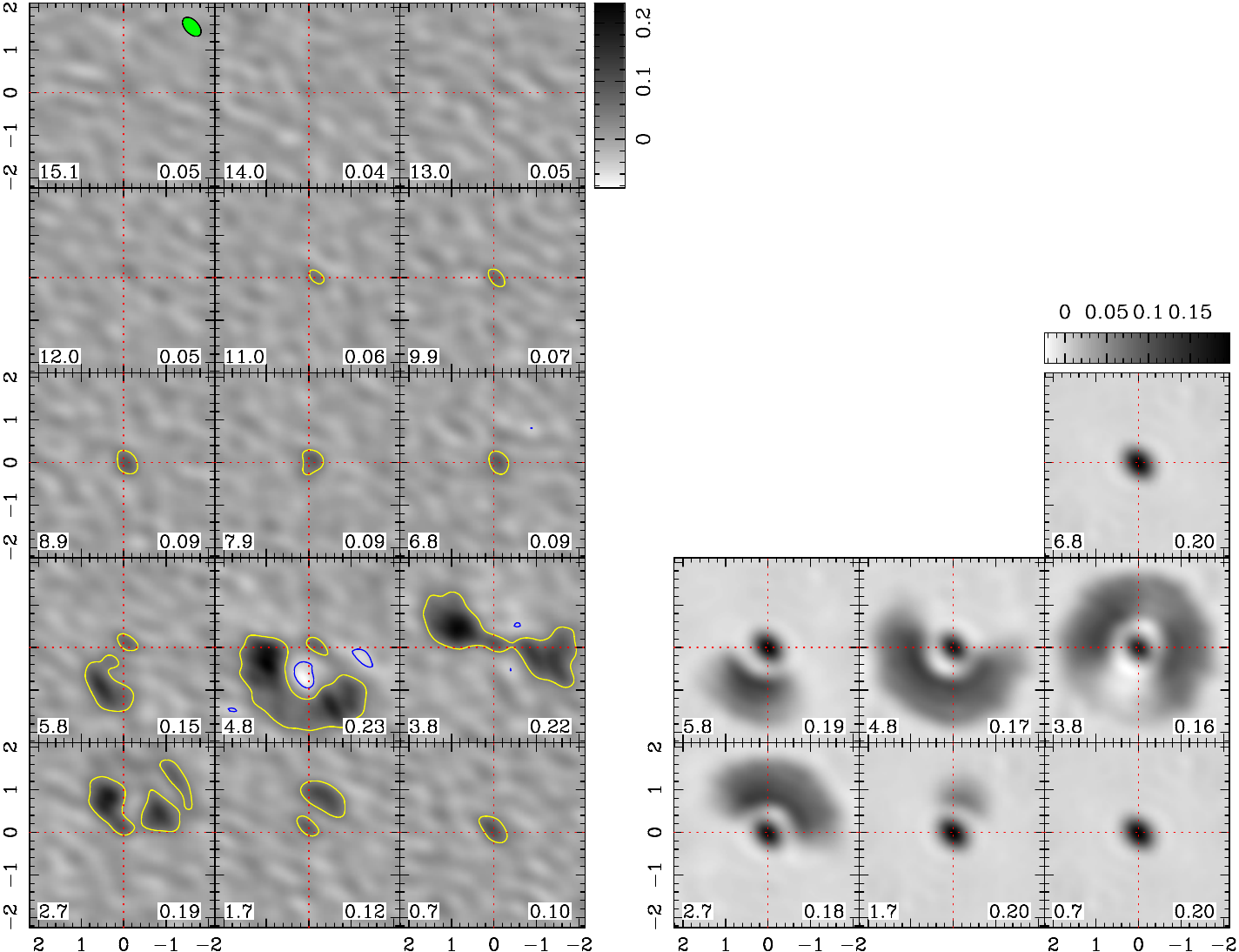}
\end{center}
\caption{{\bf Channel maps in HCO$^+$(4-3) highlight the high-velocity
    streams below 0.7~km~s$^{-1}$ and above 6.8~km~s$^{-1}$ and the
    gap-crossing filaments at 3.8~km~s$^{-1}$ (systemic velocity).}
  {\bf a)} Left column: we show specific intensity maps from CLEAN
  reconstructions in exponential grey scale for velocity bins
  corresponding to 5-channel averages, a thin yellow contour at
  4~$\sigma$ and a thin blue contour at -4~$\sigma$, where 1~$\sigma =
  12$~mJy~beam$^{-1}$ is the noise level. The LSR velocity is
  indicated at the bottom left of each image, in km~s$^{-1}$, while
  the peak specific intensity is indicated at bottom right, in
  Jy~beam$^{-1}$, with a beam of $ 0.51 \times 0.33$~arcsec$^{-2}$ (as
  in Fig.~\ref{fig:cycle0}).  Axis labels for images are in arcsec;
  North is up, East is to the left.  The cross-hairs indicate the
  origin. {\bf b)} Right column: Channel maps in HCO$^+$(4-3) from the
  fiducial model.  Same as a) but for a comparison model disk
  calculated with the LIME package, filtered in ($u,v$) coverage and
  reconstructed with CLEAN in the same way as for the ALMA
  observations. We have added a central Gaussian to the LIME model,
  which is intended to illustrate that beam elongation effects cannot
  join this central component with the outer disk. {\em Continues in
    Fig.~\ref{fig:channelsHCOplus.1}}
\label{fig:channelsHCOplus}}
\end{figure}

\newpage

\begin{figure}
\begin{center}
\includegraphics[width=0.5\textwidth,height=!]{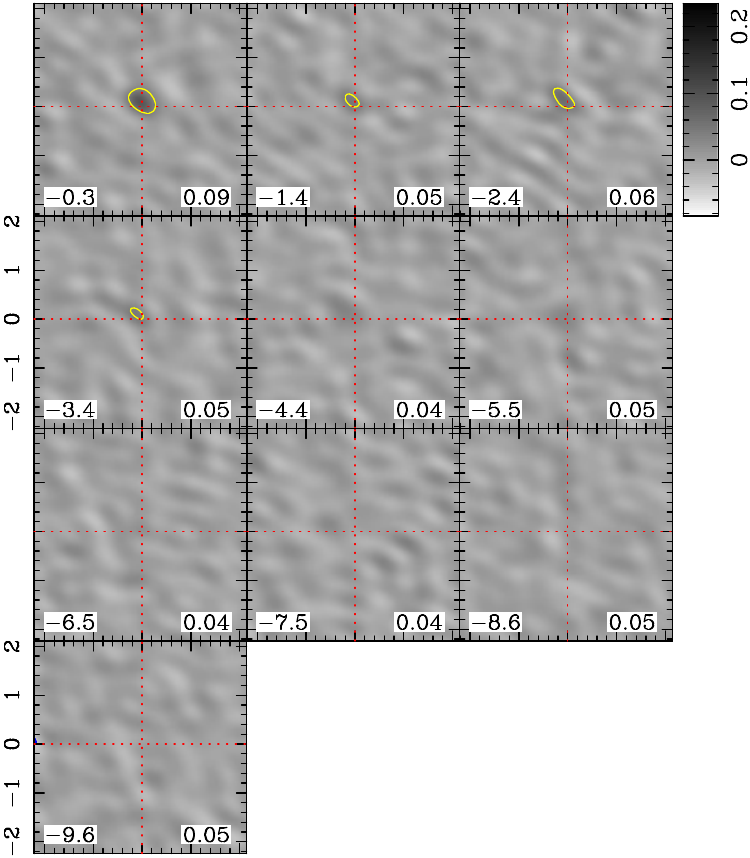}
\end{center}
\caption{{\em continues from Fig.~\ref{fig:channelsHCOplus}}
\label{fig:channelsHCOplus.1}}
\end{figure}

\newpage

%\label{fig:channelsHCOpluslime}}

\begin{figure}
\begin{center}
\includegraphics[width=0.9\textwidth,height=!]{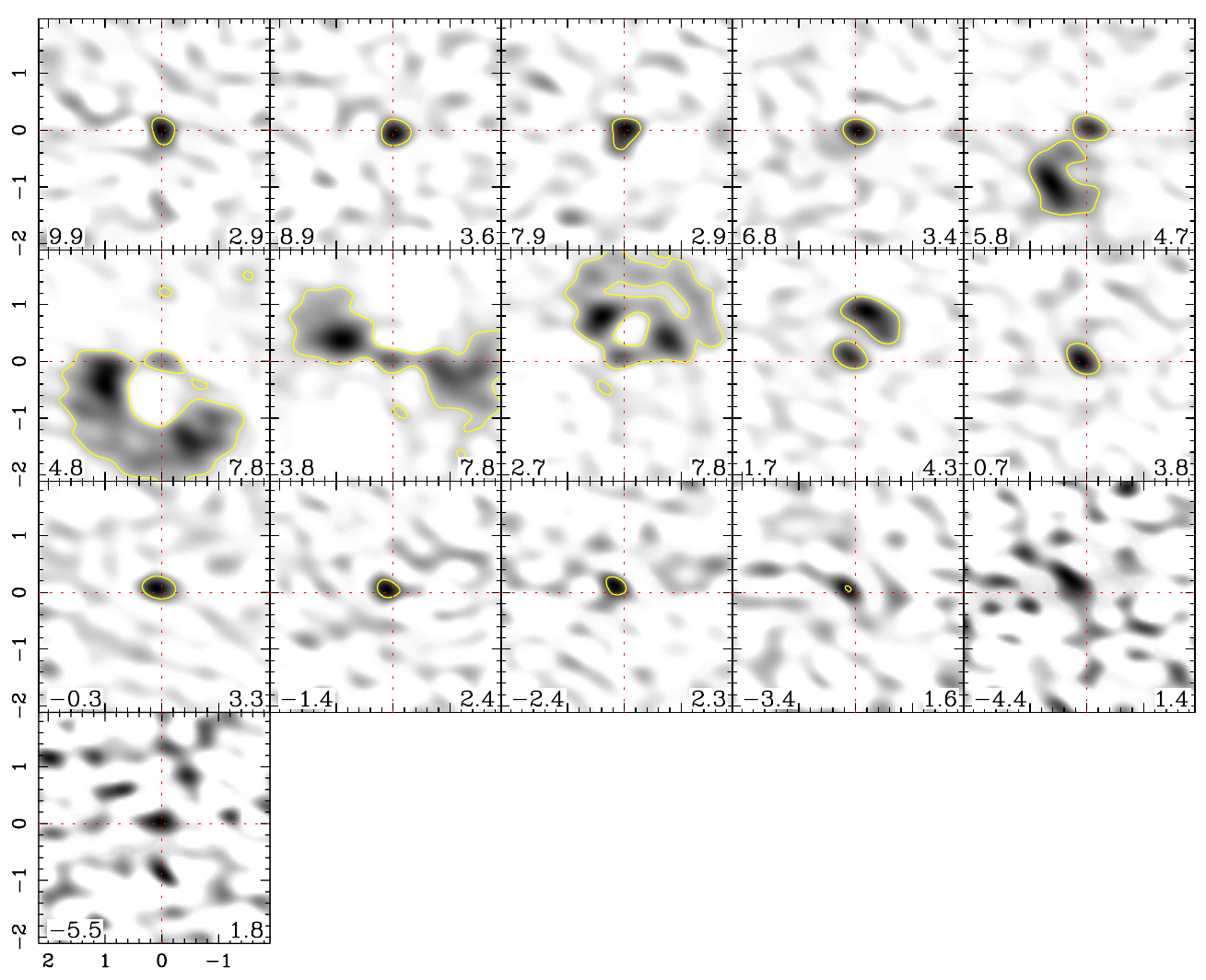}
\end{center}
\caption{{\bf Channel maps in HCO$^+$(4-3) from deconvolved models.}
  Same as Fig.~\ref{fig:channelsHCOplus} but for a deconvolved MEM
  model of the data rather than the restored CLEAN images. Peak
  intensity values are indicated in $10^{4}$MJy~sr$^{-1}$. Notice the
  gap-crossing filaments seen in systemic velocity, at
  3.8~km~s$^{-1}$. We use a reference 1~$\sigma$ value of
  3090~MJy~sr$^{-1}$. To convert into specific intensity units used
  for the restored imaged (convolved with the CLEAN beam), these
  MJy~sr$^{-1}$ must be multiplied by $4.48~10^{-6}$. 
\label{fig:channels_HCOplus_MEM}}
\end{figure}

\newpage

\begin{figure}
\begin{center}
\includegraphics[width=0.9\textwidth,height=!]{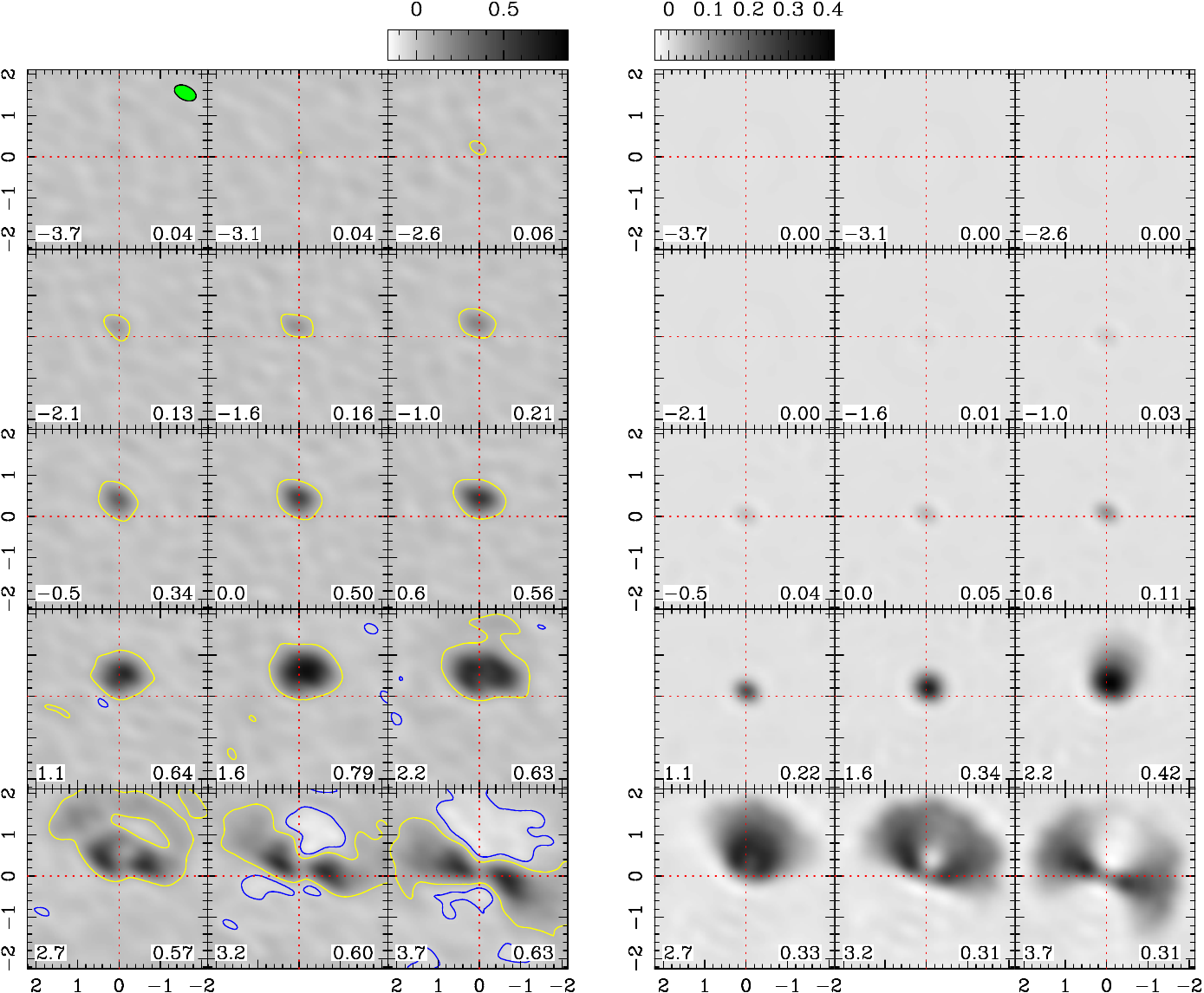}
\end{center}
\caption{{\bf Channel maps in CO(3-2).} Labels follow from
  Fig.~\ref{fig:channelsHCOplus}.  {\bf a)} Left column: we show
  specific intensity maps in grey scale for 28 velocity bins
  corresponding to 5-channel averages, with a yellow contour at
  5~$\sigma$ and a blue contour at -5~$\sigma$, where
  1~$\sigma$=0.011~Jy~beam$^{-1}$.  The LSR velocity is indicated at
  the bottom left of each image, in km~s$^{-1}$, while the peak
  specific intensity is indicated at bottom right, in Jy~beam$^{-1}$,
  with a beam of $ 0.51 \times 0.33$~arcsec$^{-2}$ (as in
  Fig.~\ref{fig:cycle0}).  Axis labels for images are in arcsec; North
  is up, East is to the left.  The cross-hairs indicate the
  origin. {\bf b)} Right column: same as Fig.~\ref{fig:channelsCO} but
  for a comparison model disk calculated with the LIME package,
  filtered in ($u,v$) coverage and reconstructed with CLEAN in the
  same way as for the ALMA observations in a). {\em Continues in
    Fig.~\ref{fig:channelsCO.1}} \label{fig:channelsCO}}
\end{figure}

\begin{figure}
\begin{center}
\includegraphics[width=0.9\textwidth,height=!]{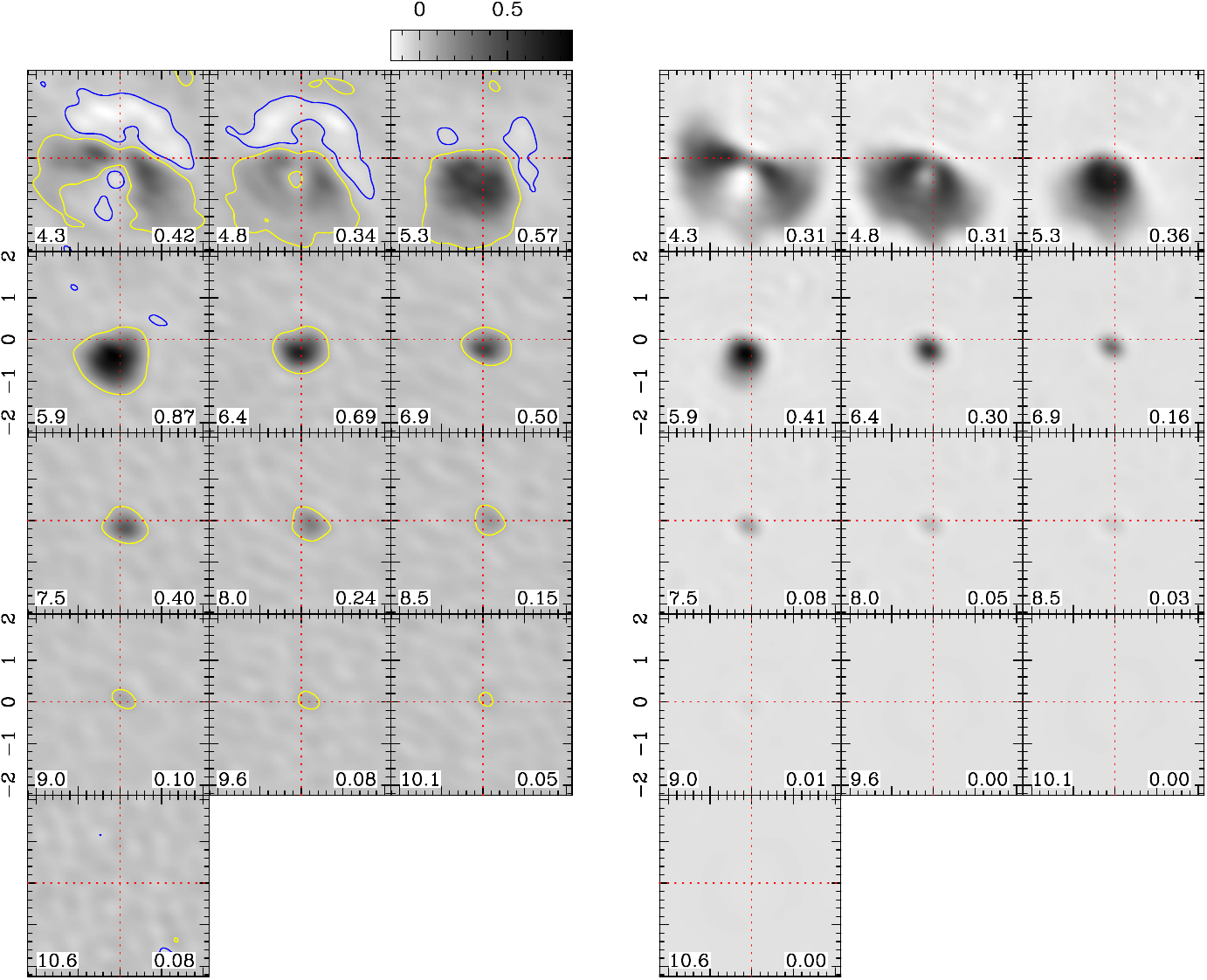}
\end{center}
\caption{{\em continues from Fig.~\ref{fig:channelsCO}}
\label{fig:channelsCO.1}}
\end{figure}

\begin{figure}
\begin{center}
\includegraphics[width=0.9\textwidth,height=!]{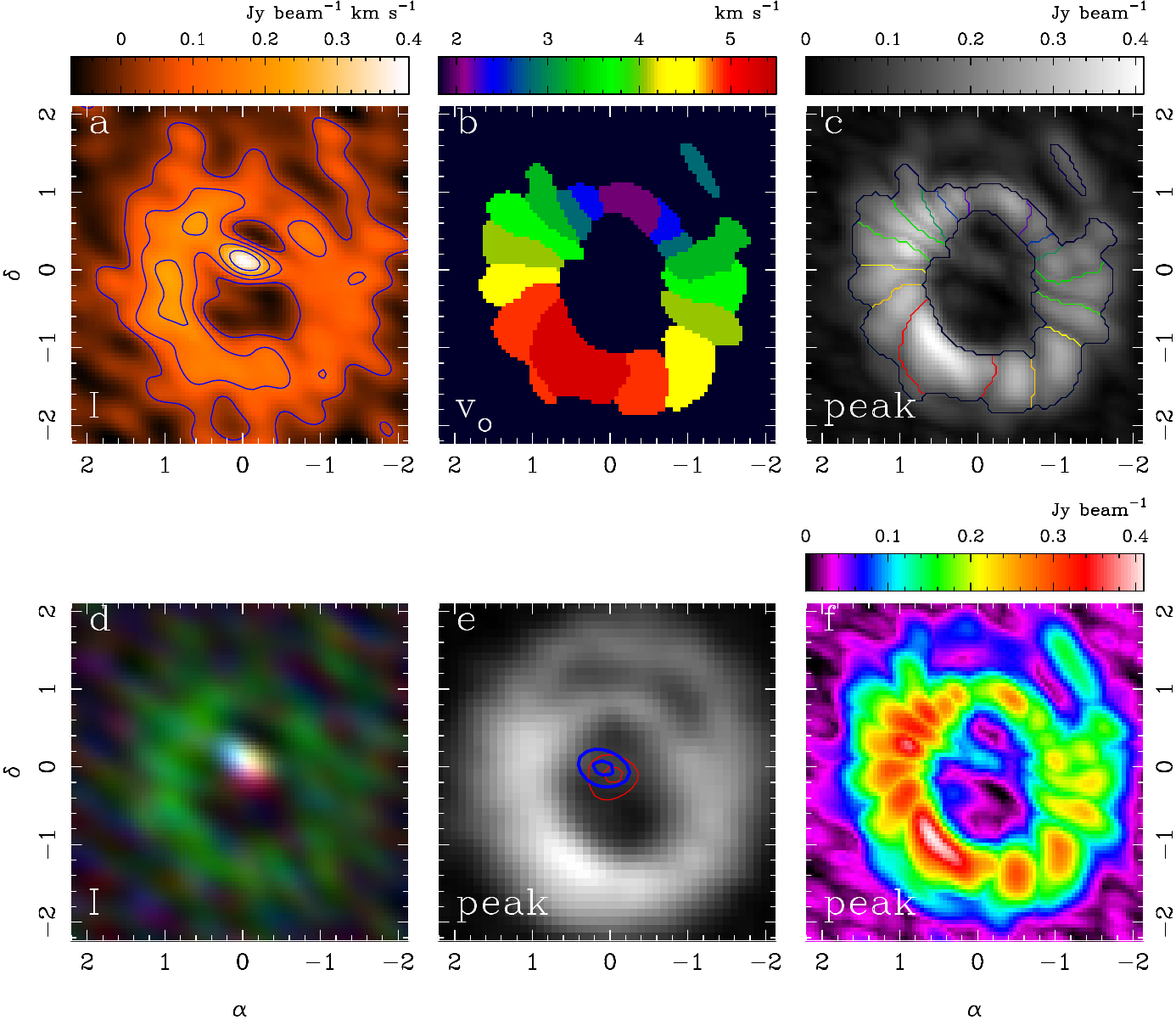}
\end{center}
\caption{{\bf HCO$^+$(4-3) maps}.  Axis labels for images are in
  arcsec; North is up, East is to the left. The origin of coordinates
  is centred on the star. The beam size is $ 0.51 \times
  0.33$~arcsec$^{-2}$ (as in Fig.~\ref{fig:cycle0}) {\bf a:} Intensity
  map, highlighting that the peak is found near the center. Contour
  levels are as in Fig.~\ref{fig:cycle0}c. {\bf b:} LSR velocity of
  the peak emission. {\bf c:} Peak intensity map, showing a whole
  ring.  The modulation along the ring that leads to a segmented
  appearance is due to the 5-channel average, as indicated by the
  exact overlap with the channel boundaries.  Countours indicate LSR
  velocity, with colour codes that match b). {\bf d:} RGB intensity
  image, with the Keplerian velocities ( $1 \le v / $km~s$^{-1} \le
  5.9$) in green, and the high velocity flows in red ($11.5 > v/
  $km~s$^{-1}>5.9$) and blue ($1 > v / $km~s$^{-1} > -4$). {\bf e:}
  Peak intensity taken on a CLEAN reconstruction of the original data,
  without channel averaging, and with subsequent smoothing by
  0.2~arcsec. As a consequence of the fine sampling, the central
  filamentary flows are not as obvious as in the 5-channel average,
  but the outer ring is less segmented.  Contours correspond to blue
  (-4.5 to +1.3~km~s$^{-1}$) and red (+5.6 to +10~km~s$^{-1}$)
  intensity, and are taken at fractions of 0.5 and 0.9 of each
  peak. {\bf f:} peak intensity, as in c), but highlighting the
  central eastern bridge in false colour and exponential scale. The
  segmented appearance is due to the 5-channel average.
 \label{fig:whcomax}}
\end{figure}

%
%\begin{figure}
%\begin{center}
%\includegraphics[width=0.5\textwidth,height=!]{panelspecMEM.pdf}
%\end{center}
%\caption{{\bf Spectra extracted on images restored from MEM models
%    recover more flux than CLEAN}. \textcolor{blue}{Probably omit this
%    because my individual restored MEM images are noisier than CLEAN
%    for now because of a problem with the weights in this .ms, and
%    because this is so spectacularly better than CLEAN that it
%    deserves another paper}.
% \label{fig:spectraMEM}}
%\end{figure}

%\begin{figure}
%\begin{center}
%\includegraphics[width=\textwidth,height=!]{hd142-Nature-hydro-v2.pdf}
%\end{center}
%\caption{{\bf Hydrodynamic simulations of the HD~142527 system.} Left:
%    Simulation considering the presence of a second stellar component
%    at 12.5~AU from the central star. Right: Simulation that includes
%    two protoplanets driving gap-crossing accretion flows.
%  \label{fig:fargo}}
%\end{figure}

\end{document}